\documentclass[a4paper,11pt]{article}
\PassOptionsToPackage{table}{xcolor}

\usepackage{jheppub} % for details on the use of the package, please
\usepackage[T1]{fontenc} % if needed

\usepackage[all]{xy}
\usepackage[percent]{overpic}
\usepackage{cancel}
\usepackage{soul}
\usepackage[normalem]{ulem}
\usepackage{slashed}
\usepackage{wrapfig}
\usepackage{tabu}
\usepackage{diagbox}
\usepackage{mathrsfs,amsmath,amssymb,amsthm,amsfonts,tikz,graphicx,accents,hyperref, color}
\usepackage{dsfont,epiolmec, latexsym, stmaryrd, comment}
\usepackage{slashed,ccaption}
\usepackage{mathrsfs, calligra}
\usepackage{leftidx}
\usepackage{import}
\usepackage{multirow}
\usepackage{amsfonts}
\usepackage{pifont}
\usepackage{tabularx}
\usepackage[utf8]{inputenc}
\usetikzlibrary{intersections,calc}
\usepackage{tikz-3dplot}
\usepackage{ifthen}
\usetikzlibrary{arrows}
\usepackage{amsmath}
\usepackage{xcolor}
\usepackage{array}
\hypersetup{ linktoc=all,
    colorlinks, linkcolor={palatinateblue},
    citecolor={brightpink}, urlcolor={amaranth}
}

\graphicspath{{Images/}}

\renewcommand{\d}[1]{\ensuremath{\operatorname{d}\!{#1}}}

\def\sideremark#1{\ifvmode\leavevmode\fi\vadjust{\vbox to0pt{\vss% the remark
 \hbox to 0pt{\hskip\hsize\hskip1em%                          will appear only
 \vbox{\hsize2cm\tiny\raggedright\pretolerance10000%          on the side
 \noindent #1\hfill}\hss}\vbox to8pt{\vfil}\vss}}}%
\DeclareSymbolFont{extraup}{U}{zavm}{m}{n}
\DeclareMathSymbol{\varheart}{\mathalpha}{extraup}{86}
\DeclareMathSymbol{\vardiamond}{\mathalpha}{extraup}{87}
\makeatletter
\renewcommand*{\@fnsymbol}[1]{\ensuremath{\ifcase#1\or \clubsuit \or \vardiamond \or \varheart\or
    \spadesuit\or \mathparagraph\or \|\or **\or \dagger\dagger
    \or \ddagger\ddagger \else\@ctrerr\fi}}
\makeatother

\definecolor{rosy}{RGB}{230,235,252}
\definecolor{myframetitle}{RGB}{90,89,170}
\definecolor{myblocktitle}{RGB}{140,185,249}
\definecolor{mytitle}{RGB}{10,80,26}

\definecolor{darkgreen}{RGB}{27,130,45}
\definecolor{darkblue}{rgb}{0,0,0.3}
\definecolor{darkred}{rgb}{0.7,0,0}

\definecolor{light gray}{RGB}{220,220,220}
\definecolor{dark purple}{RGB}{108,0,217}
\definecolor{pink}{RGB}{190,20,100}
\definecolor{orang}{RGB}{193,63,0}
\definecolor{green}{RGB}{11,98,17}
\definecolor{darkpink}{RGB}{153,0,76}
\definecolor{bluegreen}{RGB}{0,102,102}
\definecolor{greenlagan}{RGB}{0,102,0}
\definecolor{redgreen}{RGB}{102,102,0}
\definecolor{Redgreen}{RGB}{153,76,0}
\definecolor{vividviolet}{rgb}{0.62, 0.0, 1.0}
\definecolor{amaranth}{rgb}{0.9, 0.17, 0.31}
\definecolor{palatinateblue}{rgb}{0.15, 0.23, 0.89}
\definecolor{brightpink}{rgb}{1.0, 0.0, 0.5}
\definecolor{cornflowerblue}{rgb}{0.39, 0.58, 0.93}
\definecolor{deepcarminepink}{rgb}{0.94, 0.19, 0.22}
\definecolor{radicalred}{rgb}{1.0, 0.21, 0.37}
\newcommand\cnote[1]{\textcolor{red}{\bf [C:\,#1]}}

\newcommand{\be}{\begin{equation}}
\newcommand{\ee}{\end{equation}}
\newcommand{\bea}{\begin{eqnarray}}
\newcommand{\eea}{\end{eqnarray}}

\DeclareFontFamily{OT1}{rsfs}{}

\DeclareFontShape{OT1}{rsfs}{m}{n}{ <-7> rsfs5 <7-10> rsfs7 <10->rsfs10}{} 

\DeclareMathAlphabet{\mycal}{OT1}{rsfs}{m}{n}
\makeatletter \@addtoreset{equation}{section}

\begin{document}
\preprint{IPM/P-2021/--}
%\rightline{BIMSA/P-nn}

%\newcommand{\mytitle}{Super T-Witts from the horizon}
%\newcommand{\mytitle}{Symmetries Near  Null Boundaries:\\  Topologically Massive Gravity Case}
%{\tcb{Physics at The Boundary,  I) Kinematics Of Null Boundaries}}} 
%{\LARGE{Chiral Massive News Through Null Boundaries:}}
\newcommand{\mytitle}{{ {\Huge{Chiral Massive News:}} \\ \Large{Null Boundary Symmetries in  Topologically Massive Gravity}
}}
\title{\center{\textbf{\mytitle}}}

\author[a,b]{H.~Adami}
%\author[b]{, D.~Grumiller}
\author[c]{, M.M.~Sheikh-Jabbari}
\author[d]{, V.~Taghiloo}
\author[b]{, H.~Yavartanoo}
\author[e]{and C.~Zwikel}
\affiliation{$^a$ Yau Mathematical Sciences Center, Tsinghua University, Beijing 100084, China}
\affiliation{$^b$ Beijing Institute of Mathematical Sciences and Applications (BIMSA), Huairou District, Beijing 101408, P. R. China}
\affiliation{$^c$ School of Physics, Institute for Research in Fundamental
Sciences (IPM),\\ P.O.Box 19395-5531, Tehran, Iran}
%\affiliation{$^c$ The Abdus Salam ICTP, Strada Costiera 11, Trieste, Italy}
\affiliation{$^d$ Department of Physics, Institute for Advanced Studies in Basic Sciences (IASBS),
P.O. Box 45137-66731, Zanjan, Iran}
\affiliation{$^e$ Institute for Theoretical Physics, TU Wien, Wiedner Hauptstrasse 8--10/136, A-1040 Vienna, Austria}

\emailAdd{hamed.adami@bimsa.cn,  jabbari@theory.ipm.ac.ir,\\  v.taghiloo@iasbs.ac.ir,
yavar@bimsa.cn, celine.zwikel@tuwien.ac.at}

\abstract{We study surface charges on a generic null boundary in three dimensional topological massive gravity (TMG). We construct the solution phase space which involves four independent functions over the two dimensional null boundary. One of these functions corresponds to the massive chiral propagating graviton mode of TMG. The other three correspond to three surface charges of the theory, two of which can always be made integrable, while the last one can become integrable only in the absence of the chiral massive graviton flux through the null boundary. As the null boundary symmetry algebra we obtain Heisenberg $\oplus$ Virasoro algebra {with} a central charge  proportional to the gravitational Chern-Simons term of TMG. We also discuss that the flux of the chiral massive gravitons appears as the (Bondi) news through the null surface.}
\maketitle

%%%%%%%%%%%%%%%%%%%%%%%%%%%%%%%%%%%%%%%%%%%%%%%%%%%%%%%%%%%%%5
\section{Introduction}\label{sec:1}
%%%%%%%%%%%%%%%%%%%%%%%%%%%%%%%%%%%%%%%%%%%%%%%%%%%%%%%%%%%%%5
Equivalence principle is (in part) formulated  through diffeomorphism invariance of the action and the corresponding physical observables. However, this statement should be refined for spacetimes with boundaries, codimension one hypersurfaces \cite{Sheikh-Jabbari:2016lzm}. 
%\HYnote{spacelike boundaries should be excluded.} 
Among the  spacetime boundaries of  physical interest we can mention asymptotic boundaries of (A)dS or flat spacetimes or black hole horizons.  Presence of boundaries usually gives rise to new degrees of freedom which only reside at the boundary and do not propagate into the bulk, the \textit{boundary degrees of freedom} (BDoF). The  \emph{solution phase space} therefore in general contains both bulk and boundary degrees of freedom. %The former accounts for propagating, or local, degrees of freedom which cannot be removed by a symmetry transformation. 

In gauge or gravity theories BDoF can be conveniently labelled through surface charges associated with a certain sector of gauge symmetries  which preserves the boundary structure and rotate us in the solution phase space. In the existing literature this certain part of gauge symmetries is usually specified through specific falloff/boundary conditions on the fields which are appropriately chosen to describe the desired physics. It was, however, noted in \cite{Grumiller:2020vvv} that there exists a \textit{maximal boundary phase space} which can be achieved relaxing the falloff conditions and is labelled by all possible boundary preserving gauge symmetries which have finite surface charges; see also  \cite{Grumiller:2016pqb,Grumiller:2017sjh, Grumiller:2017qao} for related work.\footnote{In a series of  papers \cite{Freidel:2020xyx,Freidel:2020svx,Freidel:2020ayo} a closely related  notion of corner symmetries have been developed and studied.} For $d$ dimensional gravity cases this maximal boundary phase space is described by $d$ charges/fields which are generic functions over the boundary, a $d-1$ dimensional surface in spacetime \cite{Grumiller:2020vvv}. Then  imposing boundary or falloff conditions can be formulated through imposing second-class constraints and working with a reduced boundary phase space. 

An important property of the charges labelling the BDoF is their \textit{integrability} over the solution phase space. The basic formulation usually employed for the computation of  charges associated with the boundary preserving gauge symmetries is the covariant phase space method \cite{Lee:1990nz, Iyer:1994ys}. This formulation yields  \emph{variation of charges} in the solution phase space, and if integrable one can then define charges. Another important feature of the covariant phase space method \`a la Lee-Wald \cite{Lee:1990nz} is the ambiguities involved in the formulation. In particular, there is the so-called $Y$-ambiguity according {to} which the expression for the charge variation is ambiguous up to certain total variations over the phase space. This ambiguity may be (partially) fixed by different physical requirements. For example,  this ambiguity was used in \cite{Compere:2015bca, Compere:2015mza} to fix the central charge of the algebra of surface charges to a certain value, in \cite{Compere:2008us,Compere:2018ylh,Compere:2020lrt,Fiorucci:2020xto,Ruzziconi:2020wrb} it was used to render surface charges computed at asymptotic boundaries finite  and also to relate charges computed {in different} formulations {of gravity} \cite{DePaoli:2018erh,Oliveri:2019gvm}; {see also \cite{Margalef-Bentabol:2020teu,G.:2021qiz}}. In this work, as we will see the $Y$-ambiguity is crucially used in the charge integrability analysis.

It has been argued that the integrability is related to the absence of  flux through the boundary \cite{Wald:1999wa,Barnich:2011mi}. In \cite{Adami:2020ugu}, we made this statement more precise and conjectured that it is always possible to render the charges integrable in the absence of  ``genuine flux'' passing through that boundary, even when boundary sources are switched on. This might be accomplished by choosing a particular {field-dependent} linear combination of boundary symmetries, a \textit{field-dependent slicing} used to span the boundary phase space. This clarifies the physical notion of integrability. We note that field-dependent slicings generically change the bracket structure and hence the algebra of charges.  See \cite{Compere:2017knf,Grumiller:2019fmp, Adami:2020amw,Ciambelli:2020shy,Alessio:2020ioh} and in particular \cite{Adami:2020ugu} for examples of change of slicing and how it affects the algebra.

Recent studies of boundary degrees of freedom for gravity theories has been motivated in  pursuit of black hole microstates within the ``soft hair proposal'' \cite{Hawking:2016msc} where the near horizon symmetries and charges play a key role, e.g. see \cite{Donnay:2015abr,Donnay:2016ejv, Afshar:2016wfy, Afshar:2016kjj, Grumiller:2019fmp}. There are various types of horizons with different properties e.g. see \cite{Chrusciel:2020fql}, which regardless of the details, are usually null surfaces. In addition, null hypersurfaces  are causal boundaries of portions of the spacetimes available to the congruence of causal curves which do not intersect the null surface. Therefore, null surfaces suitably model black hole horizons and studying null boundary symmetries and charges is expected to help with the identification of black hole microstates, for a specific realization for three dimensional black holes see \cite{Afshar:2016uax, Afshar:2017okz}.

We have hence established a program of studying \textit{null boundary symmetries} and their \textit{maximal boundary phase phase}. We have explored a generic null hypersurface in  the $4d$ Einstein gravity \cite{Adami:2020amw}, $2d$ JT gravity and $3d$ Einstein-$\Lambda$ \cite{Adami:2020ugu}. Similar maximal phase space  for asymptotic boundaries in 2$d$ and 3$d$ has also been studied \cite{Ruzziconi:2020wrb}. The analysis in \cite{Adami:2020amw} was motivated by studying the phase space near horizon of  Kerr black holes,\footnote{The analysis in \cite{Adami:2020amw} does not realize the maximal phase space as described in \cite{Grumiller:2020vvv}; while we had four charges two of the charges were functions of codimension two surfaces (rather than codimension one). The analysis of the most general case will be  presented in the upcoming paper \cite{progress-1}.} while in \cite{Adami:2020ugu} we focused on  $2d$ dimensional and BTZ black holes \cite{Banados:1992gq,Banados:1992wn} where we  explicitly constructed the maximal boundary phase space. We showed that in the $2d$ case we have two charges which are functions of the $v$ coordinate along the null boundary and in the $3d$ case there are three charges which are functions over the null cylinder parametrised by $v$ and a periodic coordinate $\phi$. By an appropriate choice of slicing, we rendered the charges integrable, in accord with the conjecture made in \cite{Grumiller:2020vvv}: In the $2d$ and $3d$ cases we do not have a propagating bulk degree of freedom and there is hence no genuine flux through the null surface. The integrable slicing is not unique and there are many such slicings. In particular, we discussed there exists a  \textit{fundamental slicing}, for which the algebra of charges at any constant $v$ takes the simple form of Heisenberg algebra $\oplus$ Diff$(d-2)$ where $\hbar$ in the Heisenberg part is proportional to inverse of the corresponding Newton constant.  We should stress that the charges we find at null boundaries  are different than  {most of the} charges and algebras appeared in the literature which  are  functions of codimension two surfaces rather than codimension one surfaces.\footnote{In the AdS3/CFT2 literature, it is customary to call the boundary degrees of freedom allowed by the Brown-Henneaux boundary conditions as ``boundary gravitons''. In our maximal boundary phase space analysis, however, boundary gravitons appear as a  subsector of our BDoF.}

A natural following step in the BDoF program and a non-trivial check for the integrability conjecture is to analyze the maximal phase space of theories admitting propagating degrees of freedom. The first such example is to consider $3d$ cases. There is indeed a plethora of three dimensional beyond Einstein gravity theories which are free of ghosts, e.g. see \cite{Merbis:2014vja, Afshar:2014ffa, Ozkan:2018cxj} and have massive propagating degrees of freedom (gravitons). Among these, \textit{topologically massive gravity} (TMG) \cite{deser1982three,deser2000topologically} is special as its action involves gravitational Chern-Simons (CS) term which gives rise to massive chiral gravitons. 

Various aspects of topologically massive gravity has been extensively studied in the recent literature. While we do not have a full classification of TMG background solutions yet,  large classes of solutions have been constructed and analyzed \cite{Macias:2005pm, Chow:2009km, Chow:2009vt, chow:2019ucq, Gurses:2010sm, Gurses:2011fv, Ertl:2010dh, Deser:2009er, Garbarz:2008qn, Carlip:2008eq, Aliev:1995cf, Nutku:1993eb}. As we will review,  TMG background solutions fall into to two classes, those which have Vanishing Cotton Tensor (VCT), and those with non-vanishing Cotton tensor (NVCT). Solutions to Einstein-$\Lambda$ theory remain a solution in the presence of the CS term and constitute the VCT class. In particular, BTZ black holes are hence also solutions to TMG. Several solutions in the NVCT class has also been constructed, among them the warped solutions and other black hole solutions  \cite{Anninos:2008fx,sachs:2011xa, Bonora:2011gz, Detournay:2012ug, Compere:2009zj, Bouchareb:2007yx, Moussa:2003fc,Detournay:2015ysa}.

\subsection{Summary of the results}

In this work, we perform the null boundary symmetry analysis for an arbitrary null surface in TMG and construct the maximal boundary phase space. To this end we start with the most general metric expanded around a null surface, to be more precise a null cylinder whose axis is parametrised by ``lightcone time'' $v$ and its circle by $\phi$. We then impose TMG equations of motion, which are third order differential equations, perturbatively around the null cylinder and construct the solution phase space. The solution phase space is described by four independent functions on the null cylinder. In this case, unlike the $3d$ Einstein-$\Lambda$ theory discussed in \cite{Adami:2020ugu}, we have chiral massive gravitons which propagate in the bulk {and can pass through the null cylinder}. Three of the four functions in the 
phase space correspond to BDoF and one to the chiral graviton propagating in the bulk.

There are diffeomorphisms which keep the null boundary structure and hence act as symmetries over the solution phase space. These  are $2d$ diffeomorphisms and the Weyl scaling on the null cylinder and are hence specified by three functions over the null cylinder.
We then work through the usual formulation of covariant phase space method to compute charges associated with these symmetries.

\subsubsection*{ $\bullet$ VCT class}
There is no propagating degrees of freedom in the VCT sector and hence we are expecting integrable charges. 
Nonetheless, unlike previous cases \cite{Adami:2020ugu,Ruzziconi:2020wrb}, a choice of slicing is not sufficient to render the charges integrable. It has to be supplemented by an appropriate choice of the $Y$-ambiguity \eqref{IWamb}. This is similar to the case of asymptotic boundaries where {Lee-Wald charges  diverge and $Y$-terms are needed to render finite the charges \cite{Ruzziconi:2020wrb}.} In other words, integrability of charges can be used to (partially) fix the $Y$-ambiguity in the charges. 

In the fundamental slicing \eqref{vct-int-slicing} and \eqref{calJ-bbT-def}, the charge algebra takes the form of a Heisenberg $\oplus$ Virasoro algebra. The central charge of the Virasoro is proportional to the coefficient of the CS part; precisely it is equal to the gravitational anomaly of dual $2d$ CFT:
\begin{subequations}\label{NB-algebra-SPJ-Fourier}
    \begin{align}
      \hspace*{-8mm}  [\boldsymbol{\mathcal{S}}_n,\boldsymbol{\mathcal{S}}_m] =0,\quad 
		[\boldsymbol{\mathcal{S}}_n,\boldsymbol{\mathcal{P}}_m]&=\frac{{i}}{8G}\delta_{m+n,0}, \quad 	[\boldsymbol{\mathcal{P}}_n,\boldsymbol{\mathcal{P}}_m]=0,\\
			 [\boldsymbol{\mathcal{J}}_n,\boldsymbol{\mathcal{S}}_m]=0,&\qquad
		 [\boldsymbol{\mathcal{J}}_n,\boldsymbol{\mathcal{P}}_m]=0,\\
		 [\boldsymbol{\mathcal{J}}_n,\boldsymbol{\mathcal{J}}_m] =&(n-m) \boldsymbol{\mathcal{J}}_{n+m} +\frac{1}{4\mu G}n^3\delta_{n+m,0} \label{NB-algebra-Virasoro}
    \end{align}
\end{subequations}
where $G$ is the Newton constant and  coefficient of the CS part is $(16\pi G\mu)^{-1}$, see \eqref{TMG-action}.
%Note that the charges $\boldsymbol{\mathcal{X}}_n=\boldsymbol{\mathcal{X}}_n(v)$ are Fourier modes of generic functions of $v,\phi$. This is to be  contrasted with the usual (e.g. near horizon or asymptotic) charge analysis, where charges are only functions of codimension two surfaces, in our $3d$ case that is the circle parametrised by $\phi$. We hence call our charges as ``hyper-charges'' to distinguish them from the usual ``super-charges''.  The zero mode of $\boldsymbol{\mathcal{S}}$ is equal to the Wald entropy \cite{Wald:1993nt, Iyer:1994ys}, $\boldsymbol{\mathcal{P}}_0$ is proportional to the logarithm of the expansion of the null vector field generating the null boundary and $\boldsymbol{\mathcal{J}}_0$ is giving the angular momentum charge. We therefore, call $\boldsymbol{\mathcal{S}}_n(v)$ \emph{hyper-entropy}, $\boldsymbol{\mathcal{P}}_n(v)$ \emph{hyper-expansion} and $\boldsymbol{\mathcal{J}}_n(v)$ \emph{hyper-rotation}.
 Note that the charges $\boldsymbol{\mathcal{X}}_n=\boldsymbol{\mathcal{X}}_n(v)$ are Fourier modes of generic functions of $v,\phi$. This is to be  contrasted with the usual (e.g. near horizon or asymptotic) charge analysis, where time dependence of charges is fixed. %We hence call our charges as ``hyper-charges'' to distinguish them from the usual ``super-charges''. 
 The zero mode of $\boldsymbol{\mathcal{S}}$ is equal to the Wald entropy \cite{Wald:1993nt, Iyer:1994ys}, $\boldsymbol{\mathcal{P}}_0$ is proportional to the logarithm of the expansion of the null vector field generating the null boundary and $\boldsymbol{\mathcal{J}}_0$ is giving the angular momentum charge. We therefore, call $\boldsymbol{\mathcal{S}}_n(v)$ \emph{entropy aspect charge}, $\boldsymbol{\mathcal{P}}_n(v)$ \emph{expansion aspect charge} and $\boldsymbol{\mathcal{J}}_n(v)$ \emph{angular momentum aspect charge}. %To simplify the terminology, whenever there is no confusion, we drop the ``charge'' and call these charges simply as entropy aspect, expansion aspect and angular momentum aspect.

Finally, we consider non-expanding backgrounds. The expansion aspect charge vanishes and the algebra reduces to the one obtained in \cite{Grumiller:2019fmp} as the symmetries near a generic Killing horizon, with the addition that our charges are still $v$-dependent.\footnote{Note that while all Killing horizons are non-expanding null surfaces, the converse is not necessarily true.} As a particular example we study charges associated with BTZ black holes in TMG.

\subsubsection*{$\bullet$ NVCT class}
In the NVCT case, there is in general a non-zero flux due to massive gravitons passing through the null boundary. Hence the charges are not expected to be integrable. As we are dealing with intrinsically non-integrable charges, we use the \textit{modified bracket} \cite{Barnich:2011mi}  \eqref{BT-Bracket-01} to extract out the integrable part of the charge variations. The remainder is then associated with the flux passing through the surface  \eqref{charge-NVC-case-non-integrable part}.

Our charge analysis involves three steps: (1)  finding appropriate slicing of the phase space; (2) fixing the $Y$-ambiguity and (3) separating the integrable part of charge and identifying the flux. These three steps should be worked through in accord with each other. The first two steps are    similar to the VCT case while the last step is particular to the NVCT case. 

Going through these steps and after a tedious and technical analysis, we find that the charge algebra for the generic NVCT case is a Heisenberg $\oplus$ Virasoro algebra where the central charge is the gravitational anomaly of the theory. It is the same algebra \eqref{NB-algebra-SPJ-Fourier} as for the VCT but with the modified bracket.\footnote{Note that this central term is field-independent. This may be contrasted with the case of $4d$ asymptotic symmetry analysis where the central extension is field dependent \cite{Barnich:2011mi}.}  This ensures that the limit of vanishing flux is consistent with the VCT results. Moreover, the flux \eqref{charge-NVC-case-non-integrable part}  is fully sourced by the expansion of the null boundary and is associated to a symmetry perpendicular to the null surface. 

We also discuss the vanishing expansion NVCT cases. In these cases the flux vanishes and the charges become integrable. Nonetheless, unlike the VCT counterpart, there are still three independent charges, in particular we still have  an expansion aspect charge. To the best of our knowledge, TMG solutions in this class was not discussed in the literature. 

Finally, we study charges of axisymmetric TMG solutions, including the warped black holes \cite{Anninos:2008fx},  and show that they have a vanishing expansion aspect charge and fall in the same category as non-expanding backgrounds in the VCT class.

\subsection{Organization of the paper}

In section \ref{sec:2}, we review some basic facts about TMG, its action, equations of motion and how to compute the surface charges for the theory.  In section \ref{sec:NNB-metric}, we present a general near null surface expansion of metric and show that the Diff$(C_2) \oplus$ Weyl$(C_2)$ algebra, a part of the $3d$ diffeomorphisms, preserves the null boundary which is the null cylinder $C_2$. This section and its results do not depend on theory and is hence in common with those in \cite{Adami:2020ugu}. In section \ref{sec:VCT}, we study the VCT class of solutions and the associated surface charges. The latter involves fixing the $Y$-ambiguity and finding the appropriate (fundamental) slicing. As special cases we study  null boundaries with zero expansion
as well as charges for the BTZ background. In section \ref{sec:NVCT}, we repeat the same analysis but now for the generic class of NVCT. This case  importantly features non-vanishing flux and non-integrable charges, and hence involves a whole lot more technical issues, especially the modified bracket. We also discuss the special non-expanding NVCT case where the charges are integrable and there is no flux.   In section \ref{appen:axisymmetric-TMG-soln}, we construct all axisymmetric TMG solutions, in an expansion around a null surface, these involve both VCT and NVCT solutions. Last section includes our concluding remarks and some future directions.

%%%%%%%%%%%%%%%%%%%%%%%%%%%%%%%%%%%%%%%%%%%%%%%%%%%%%%%%%%%%%5
\section{Topologically massive gravity and its surface charges}\label{sec:2}
%%%%%%%%%%%%%%%%%%%%%%%%%%%%%%%%%%%%%%%%%%%%%%%%%%%%%%%%%%%%%5
Topologically massive gravity (TMG), with negative cosmological constant $\Lambda= -1/\ell^2$, is described by the action \cite{deser1982three,deser2000topologically},
\begin{equation}\label{TMG-action}
S[g]=\frac{1}{16\pi G}\int \mathrm{d}^{3}x\ L[g],\qquad L[g]:=\sqrt{-g} \left(R+\frac{2}{\ell^2}+\frac{1}{\mu}L_{\text{\tiny CS}} \right)
\end{equation}
where $R$ is Ricci scalar and $L_{\text{\tiny CS}}$ is the gravitational Chern-Simons term,
\begin{equation}
L_{\text{\tiny CS}}= \frac{1}{2}\epsilon^{\mu\nu\rho}\left(\Gamma^{\alpha}_{\mu\beta}\partial_{\nu}\Gamma^{\beta}_{\rho\alpha}+\frac{2}{3}\Gamma^{\alpha}_{\mu\beta}\Gamma^{\beta}_{\nu\gamma}\Gamma^{\gamma}_{\rho\alpha}\right).
\end{equation}
with $\epsilon_{\mu\nu\lambda}$ being the Levi-Civita tensor which in our conventions  $\sqrt{-g}\epsilon^{vr\phi}=1$, and $\Gamma^\alpha_{\mu\nu}$ is the Christoffel symbol. This action has three parameters of dimension length, $G, \ell$ and the Chern-Simons coupling $1/\mu$. One may hence construct two dimensionless constants out of their ratios. These two may be taken to be $\ell/G$ and $\mu\ell$.

TMG action is constructed out of the metric, a covariant quantity under diffeomorphisms $\delta_\xi g_{\mu \nu}=\mathcal{L}_\xi g_{\mu \nu}$ where $\mathcal{L}_\xi$ denotes the Lie derivative along vector field $\xi^\mu$, and  the connection $\Gamma^{\alpha}_{\mu \nu}$, a non-covariant quantity,
\begin{equation}
    \delta_\xi \Gamma^{\alpha}_{\mu \nu}= \mathcal{L}_\xi \Gamma^{\alpha}_{\mu \nu}+ \partial_\mu \partial_\nu \xi^\alpha \, , 
\end{equation}
where the first term in the above is defined as the Lie derivative of $\Gamma^\alpha_{\mu\nu}$ treated as
a usual tensor with the same index structure. 
Due to the presence of CS term which involves the connection, the TMG Lagrangian \eqref{TMG-action} is not diffeomorphism invariant. Variation of the Lagrangian density induced by the diffeomorphism generated by vector field $\xi$ is
\begin{equation}
    \begin{split}
    \delta_\xi L = & \mathcal{L}_{\xi} L[g] + \partial_\mu \Xi^\mu_{\xi} [g]\\
    = & \partial_\mu \left( \xi^\mu L + \Xi^\mu_{\xi}  \right) \,
    \end{split}
\end{equation}
where
\begin{equation}\label{Xi}
    \Xi^\mu_{\xi} [g]= \frac{\sqrt{-g}}{{32} \pi G \mu} \epsilon^{\mu \nu \lambda} \partial_\nu \Gamma^\alpha_{\lambda\beta} \partial_\alpha \xi^\beta \, ,
\end{equation}
encodes the non-invariant part.

Holographic aspects of asymptotically locally (warped) AdS$_3$ solutions of TMG has been analyzed in \cite{Grumiller:2017otl, Anninos:2010pm, Blagojevic:2009ek, Henneaux:2011hv, Merbis:2014vja, Detournay:2012pc,Skenderis:2009nt}. Moreover, asymptotic symmetry analysis of the asymptotically AdS$_3$ solutions with Brown-Henneaux boundary conditions \cite{Brown:1986nw} has also been performed \cite{Grumiller:2013at, Henneaux:2010fy, Henneaux:2009pw, Compere:2008cv, Hotta:2008yq}. It has been shown that we get two Virasoro algebras at different left and right central charges $c_L, c_R$, where $c_L+c_R=3\ell/G$ is twice the usual Brown-Henneaux central charge  and $c_L-c_R=\frac{3}{\mu G}$  gives the gravitational anomaly of the presumed dual $2d$ CFT, which corresponds to the non conservation of the dual stress energy \cite{Solodukhin:2005ah,Kraus:2005zm, Kraus:2006wn}. This latter is encoded in the diffeomorphism non-invariance of the TMG Lagrangian, as discussed above.

\subsection{TMG equations of motion}\label{sec:2.1}
The first-order variation of the Lagrangian $L$ is\footnote{In three dimensions we have the useful identity $V^{[\mu}\epsilon^{\nu]\rho\lambda}=\frac12(\epsilon^{\mu\nu\rho}V^\lambda-\epsilon^{\mu\nu\lambda} V^\rho )$}
\begin{equation}\label{Variation-L}
    \delta L = E^{\mu \nu}[g] \delta g_{\mu \nu}+ \partial_\mu \Theta^{\mu}[\delta g; g] \, ,
\end{equation}
where 
\begin{equation}\label{TMG-eom}
E^{\mu \nu}[g]:= - \frac{\sqrt{-g}}{16\pi G} \left( G^{\mu\nu}- \frac{1}{\ell^2} g^{\mu\nu}+\frac{1}{\mu}C^{\mu\nu} \right)=0,
\end{equation}
gives the equation of motion and
\begin{equation}\label{symplpot}
    \Theta^{\mu}[\delta g; g]=\frac{\sqrt{-g}}{16\pi G}\left[2\nabla^{[\alpha} \left(g^{\mu]\beta}\delta g_{\alpha\beta}\right)+\frac{1}{{2}\mu}\epsilon^{\lambda\mu\nu}\left(\Gamma^{\alpha}_{\lambda\beta}\delta \Gamma^{\beta}_{\nu\alpha}-2R^{\alpha}_{\lambda}\delta g_{\alpha\nu}\right) \right]\,,
\end{equation}
is the Lee-Wald symplectic potential. In \eqref{TMG-eom}, $G_{\mu\nu}$ is Einstein tensor and 
\begin{equation}\label{Cotton-tensor}
    C^{\mu\nu}:=\epsilon^{\alpha\beta\mu}\nabla_{\alpha}S_\beta^\nu, \qquad  S_{\mu \nu}= R_{\mu \nu}-\frac{1}{4} g_{\mu \nu} R= G_{\mu \nu}+\frac{1}{4} g_{\mu \nu} R, \end{equation}
$C^{\mu\nu}$ is the Cotton tensor and  $S_{\mu \nu}$ is the ${3d}$ Schouten tensor. The Cotton tensor is trace-less and divergence-free and hence the equations of motion \eqref{TMG-eom} imply $R=-\frac{6}{\ell^2}$. The Cotton tensor in $3d$ is a substitute for the Weyl tensor. It is conformally invariant and its vanishing is equivalent to conformal flatness. 

Equations of motion are a system of third order partial differential equations which may also be  written as
\begin{equation}\label{extra-condition}
    \mathcal{E}^{\mu}_{\nu} := %{}^{(M)}
    \mathcal{D}^{\mu}{}_{\beta} \, \mathcal{T}^{\beta}{}_{\nu}=0\,, 
\end{equation}
where
\begin{equation}\label{GRE}
    \mathcal{T}_{\mu \nu}:=R_{\mu \nu}+ \frac{2}{\ell^2} g_{\mu \nu}\, ,\qquad 
    \mathcal{D}^{\mu}{}_{\nu} = \delta^{\mu}{}_{\nu}+ \frac{1}{\mu} \epsilon^{\mu \alpha}{}_{\nu} \nabla_{\alpha}\, .
\end{equation}
One can simply check that on-shell $\mathcal{T}= \mathcal{T}^{\mu}_{\mu}=0,  \nabla_\nu \mathcal{T}^{\mu \nu}=0$. 

Note that while equations of motion $\mathcal{E}_{\mu\nu}=0$ are covariant, the symplectic potential is not a covariant vector: 
\begin{equation}\label{001}
    \delta_\xi \Theta^\mu[\delta g ; g]= \mathcal{L}_\xi \Theta^\mu[\delta g ; g] + \Pi^\mu_\xi[\delta g ; g] \, ,
\end{equation}
where
\begin{equation}
    \Pi^\mu_\xi[\delta g ; g] := \frac{\sqrt{-g}}{32 \pi G\mu} \epsilon^{\mu \nu \lambda} \partial_\nu \delta\Gamma^\alpha_{\lambda\beta} \partial_\alpha \xi^\beta - \partial_\nu \left( \frac{\sqrt{-g}}{32 \pi G\mu}\epsilon^{\mu \nu \lambda} \delta\Gamma^\alpha_{\lambda\beta} \partial_\alpha \xi^\beta\right)  \, .
\end{equation}

\subsection{TMG surface charges}\label{sec:2.2}

Being diffeomorphism non-invariant, one should revisit the usual Noether-Wald method \cite{Iyer:1994ys} for computing surface charges, see e.g. \cite{Tachikawa:2006sz}. Assuming that the variation \eqref{Variation-L} is induced by an infinitesimal transformation generated by $\xi$, one can define an on-shell conserved current,
\begin{equation}\label{current}
    J^\mu_\xi[g] := \Theta^\mu [\delta_\xi g ; g]- \xi^\mu L[g] -\Xi^\mu_{\xi}[g] \, .
\end{equation}
By virtue of Poincar\'e lemma, one finds
\begin{equation}
    J^\mu_\xi[g] \approx \partial_\nu K_\xi^{\mu \nu}[g] \, ,
\end{equation}
where $\approx$ denotes on-shell equality and 
\begin{equation}
    K_\xi^{\mu \nu}[g] = -\frac{\sqrt{-g}}{8 \pi G } \left[  \nabla^{[\mu}\xi^{\nu]}+ \frac{1}{\mu} \epsilon^{\mu \nu \lambda}\left( S_{\lambda \alpha }\xi^\alpha - \frac{1}{4} \Gamma^{\alpha}_{\lambda \beta} \nabla_\alpha \xi^\beta\right)\right] \, 
\end{equation}
is the Noether potential. 
Using \eqref{current},  the Lee-Wald symplectic is current,
\begin{equation}
    \begin{split}
        \omega_{\text{\tiny LW}}^\mu [\delta g, \delta_\xi g ; g] :=\ & \delta \Theta^\mu[\delta_\xi g ; g]- \delta_\xi \Theta^\mu[\delta g ; g]-\Theta^\mu[\delta_{\delta \xi} g ; g]\\
        \approx\ & \partial_\nu \left(\delta K_\xi^{\mu \nu}[g]-K_{\delta\xi}^{\mu \nu}[g]+2 \xi^{[\mu} \Theta^{\nu]}[\delta g ; g]+\Sigma^{\mu \nu}_\xi [\delta g ; g]\right) 
    \end{split}
\end{equation}
where we used the identity
\begin{equation}
\begin{split}
\delta \Xi_\xi^\mu[g] - \Xi_{\delta\xi}^\mu[g]&-\Pi_\xi^\mu[\delta g ; g] = \partial_\nu \Sigma^{\mu \nu}_\xi [\delta g ; g] \, ,\\  
\Sigma^{\mu \nu}_\xi [\delta g ; g] &:= \frac{\sqrt{-g}}{32 \pi G\mu}\epsilon^{\mu \nu \lambda} \delta\Gamma^\alpha_{\lambda\beta} \partial_\alpha \xi^\beta \, .
\end{split}
\end{equation}

The surface charge variation is defined as $\slashed{\delta} Q_{\xi} := \int_\Sigma \omega^\mu [\delta g, \delta_\xi g ; g] \d x_\mu$ and hence the surface charge variation associated to the vector field $\xi$ is \cite{Kim:2013cor}
\begin{equation}\label{TMG-surface-charge}
        \slashed{\delta} Q_{\text{\tiny LW}}(\xi) := \oint_{\partial \Sigma} \Big(\mathcal{Q}^{\mu \nu}_{\text{\tiny GR}}+ \frac{1}{\mu} \mathcal{Q}^{\mu \nu}_{\text{\tiny CS}} \Big) \d x_{\mu \nu}
\end{equation}
where $\mathcal{Q}^{\mu \nu}_{\text{\tiny GR}}$ and $\mathcal{Q}^{\mu \nu}_{\text{\tiny CS}}$ are respectively contributions of Einstein-Hilbert and the CS parts,
%\begin{eqnarray}
\begin{align}
    \mathcal{Q}^{\mu \nu}_{\text{\tiny GR}} &=\frac{\sqrt{-g}}{8 \pi G}\, \Big( h^{\lambda [ \mu} \nabla _{\lambda} \xi^{\nu]} - \xi^{\lambda} \nabla^{[\mu} h^{\nu]}_{\lambda} - \frac{1}{2} h \nabla ^{[\mu} \xi^{\nu]} + \xi^{[\mu} \nabla _{\lambda} h^{\nu] \lambda} - \xi^{[\mu} \nabla^{\nu]}h \Big),\label{GR-contribution}\\
    \mathcal{Q}^{\mu \nu}_{\text{\tiny CS}} &= \frac{\sqrt{-g}}{8 \pi G}\, \epsilon^{\mu \nu \lambda} \left(\frac{1}{2} \delta \Gamma^{\alpha}_{\lambda \beta} \nabla_\alpha \xi^\beta- \delta S_{\lambda \alpha} \xi^\alpha - \xi^\beta S^\alpha_{[\beta} h_{\lambda]\alpha} \right)\,.\label{CS-contribution}
\end{align}
%\end{eqnarray}
Here $h_{\mu \nu}= \delta g_{\mu \nu}$  denotes metric perturbations, $h= g^{\mu \nu}h_{\mu \nu}$ and the integration is over $\partial\Sigma$ which is a one-dimensional compact spacelike surface. 

\paragraph{$Y$-ambiguity.} One can readily observe that the above  definitions leave room for a freedom, an ambiguity, in the choice of the symplectic potential. {Without changing \eqref{Variation-L}, }One can shift the Lee-Wald symplectic potential as, 
\begin{equation}\label{IWamb}
    \Theta^{\mu}[\delta g ; g] \rightarrow \Theta^{\mu}_{\text{\tiny Y}}[\delta g ; g] = \Theta^{\mu}[\delta g ; g]+ \partial_\nu Y^{\mu \nu}[\delta g ; g] \,.
\end{equation}
This $Y$-ambiguity, however, affects the expression for charge variation:
\begin{equation} \label{chargeYamb}
    \slashed{\delta} Q(\xi) = \slashed{\delta} Q_{\text{\tiny LW}}(\xi) + \int_{\partial\Sigma} \d x_{\mu \nu} \mathcal{Y}^{\mu \nu}[\delta g, \delta_\xi g ; g] \, ,
\end{equation}
where
\begin{equation}
    \mathcal{Y}^{\mu \nu}[\delta g, \delta_\xi g ; g]:= \delta Y^{\mu \nu}[\delta_\xi g ; g]- \delta_\xi Y^{\mu \nu}[\delta g ; g] - Y^{\mu \nu}[\delta_{\delta\xi} g ; g] \, .
\end{equation}
As we will see later, fixing this ambiguity is a crucial part of charge analysis for the TMG theory. 

{We would like to note that while $\Gamma^\mu_{\alpha\beta}$ is not a covariant object, its variation $\delta\Gamma^\mu_{\alpha\beta}$ is. As a consequence of covariant phase space formalism in which symplectic current is defined as an anti-symmetric bi-linear in variation of fields and despite diffeomorphism non-invariance of the TMG action,  as explicitly seen,   \eqref{GR-contribution}, \eqref{CS-contribution} are covariant.  Addition of a $Y$-term, as the above general discussion shows, does not change this fact.}

%%%%%%%%%%%%%%%%%%%%%%%%%%%%%%%%%%%%%%%%%%%%%%%%%%%%%%%%%%%%%%%%%%%
\section{Near null boundary metric}\label{sec:NNB-metric}
%%%%%%%%%%%%%%%%%%%%%%%%%%%%%%%%%%%%%%%%%%%%%%%%%%%%%%%%%%%%%%%%%%%

{In this section we adopt an appropriate coordinate system to describe null boundaries and derive the symmetry preserving the boundary structure. This will be used to compute the surface charges in section \ref{sec:VCT} and \ref{sec:NVCT}. }
%%%%%%%%%%%%%%%%%%%%%%%%%%%%%%%%%%%%%%%%%%%%%%%%%%%
\subsection{Metric expansion}
%%%%%%%%%%%%%%%%%%%%%%%%%%%%%%%%%%%%%%%%%%%%%%%%%%%

Consider a null surface ${\cal N}$ and choose the coordinate system such that it sits at $r=0$. 
Here, we restrict ourselves to the family of solutions in which metric components  are smooth and analytic functions in $r$ near the null surface. Near ${\cal N}$ the metric takes  the form
\begin{equation}\label{line-element-001}
    \d s^2= -V \d v^2 + 2 \eta \d v \d r+ {\cal R}^2 \left( \d \phi + U \d v \right)^2\, , 
\end{equation}
with $\eta=\eta(v,\phi)$ and
\begin{subequations}\label{r-expansion-3d'}
    \begin{align}
       V(v,r,\phi) =&  \,r V_1(v,\phi) +r^2 V_2(v,\phi)+\mathcal{O}(r^3) \\
       U(v,r,\phi) =& \,\mathcal{U}(v,\phi) + r U_1(v,\phi) %+r^2 U_2(v,\phi)
       +\mathcal{O}(r^2) \\
       {\cal R}^2(v,r,\phi) =&\, \Omega(v,\phi)^2 + r R_1(v,\phi)% +r^2 h_2(v,\phi) 
       +\mathcal{O}(r^2)
    \end{align}
\end{subequations}
where we keep the orders relevant to the charge analysis.
The induced geometry on $\mathcal{N}$ reads
\begin{equation}\label{metric-on-N}
   \d s^2_{\text{\tiny $\mathcal{N}$}}= \Omega^2 \left( \d \phi + \mathcal{U} \d v \right)^2\, .
\end{equation}

To decompose  the bulk metric adapted to the study of null hypersurfaces, it is convenient to define two null vector fields $l^\mu, n^\mu$ ($l^2=n^2=0$) such that $l\cdot n=-1$,  $l^\mu$ is outward pointing and $n^\mu$ inward pointing
\begin{equation}\label{gennullbndryl}
%\begin{align}
    l :=  l_\mu \d x^\mu= -\frac{1}{2} V \d v  + \eta \d r , \qquad
    n :=  n_\mu \d x^\mu= -\d v \, .
%\end{align}
\end{equation}
In terms of these,
\begin{equation}\label{g-q-nl}
    g_{\mu \nu}= q_{\mu \nu}- l_\mu n_\nu - l_\nu n_\mu \, ,\qquad q_{\mu \nu} l^\mu=q_{\mu \nu} n^\mu=0.
\end{equation}
where $q_{\mu\nu} \d x^\mu \d x^\nu={\cal R}^2 \left( \d \phi + U \d v \right)^2$.  The vector field $l^\mu \partial_\mu = \partial_v - \mathcal{U} \partial_\phi + \mathcal{O}(r)$ generates the null surface and $\mathcal{U}$ can be treated as the angular velocity aspect of the given null surface. The extrinsic geometry is encoded in the expansions of $l^\mu, n^\nu$ on $\mathcal{N}$,\footnote{Since $l^\mu$ is outward pointing, $\Theta_{_{l}}\geq \Theta_{_{n}}$ or $\chi+\tau\geq 0$.}
\begin{subequations}
\begin{align}
\Theta_{_{l}} := (q_{\mu \nu} \nabla ^\mu l^\nu) \big|_{r=0} = \frac{\chi}{\Omega} \,, \qquad 
%\text{ with } 
\chi&:=\partial_v \Omega - \partial_\phi \left( {\Omega}\,\mathcal{U}\right) \label{Theta-chi-def}\,,\\
\Theta_{_{n}} :=q^{\mu\nu}\nabla_\mu n_\nu\big|_{r=0}= {-}\frac \tau\Omega , \qquad \tau&:=\frac{R_1}{2\eta\Omega} \label{Theta-tau-def}
\end{align}
\end{subequations}
and in the twist field
\begin{equation}
    \omega := -(q_\phi{}^\nu n_\lambda \nabla_\nu l^\lambda) \big|_{r=0}= \frac{1}{2} \left( \frac{\Upsilon}{\Omega}+ \frac{\partial_\phi \eta}{\eta}\right) \,, \qquad %\text{ with }
    \Upsilon:= - \frac{\Omega^3 \, U_1}{\eta}\, .
\end{equation}
For later convenience we define the non-affinity  $l \cdot \nabla l^{\mu}:= \kappa\, l^{\mu}$ on $\mathcal{N}$
\begin{equation}\label{kappa-l}
   \kappa =  \frac{V_1}{2\eta} + \frac{\partial_v \eta}{\eta} -\frac{\mathcal{U}\, \partial_\phi \eta}{\eta} \,,
\end{equation}
and
\begin{align}\label{calP-def}
     \mathcal{P}:= &\ln{\left( \frac{\eta}{\chi^2}\right)}\, , \\
    \varpi:= &\frac{\Omega^2}{\ell^2}+ \left(  \frac{\Upsilon}{2\Omega}\right)^2 + \left(  \frac{\partial_\phi \eta}{2\eta}\right)^2 + \frac32\left(  \frac{\partial_\phi \Omega}{\Omega}\right)^2 -\frac{\partial^2_\phi \Omega}{\Omega}-2 \, \tau\,\chi - 2\tau \, \Omega \, \partial_\phi \mathcal{U}
  -\mathcal{U}\partial_\phi (\tau \,\Omega) \,. 
 \end{align}
Note that $\Omega, {\cal P}, \Upsilon$ are the combinations which appeared in the charge expressions in the Einstein gravity case analyzed in \cite{Adami:2020ugu}.

\paragraph{Curvature components.}
In the rest of the work, the on-shell divergence-free and traceless tensor $\mathcal{T}_{\mu \nu}$ \eqref{GRE} will be of great relevance. 
The components $\mathcal{T}_{ll}=l^\mu l^\nu \mathcal{T}_{\mu\nu}$, $\mathcal{T}_{l\phi}=l^\mu  \mathcal{T}_{\mu\phi}$ and $\mathcal{T}_{\phi\phi}= \mathcal{T}_{\phi\phi}$ computed at $r=0$ are given by 
\begin{subequations}\label{Tmunu-components}
\begin{align}\label{Tll}
 & \mathcal{T}_{ll}= -\frac1{\Omega}\left( \partial_v \chi - \partial_\phi (\mathcal{U} \chi)-\kappa \chi \right)\, \\
   & \mathcal{T}_{l\phi}= \partial_v \omega - \partial_\phi (\mathcal{U} \omega)+ \frac{\chi \, \omega}{\Omega} - \partial_\phi \kappa \\
    &\mathcal{T}_{\phi\phi}=-2\Omega\left(
   \partial_v \tau - \partial_\phi (\mathcal{U} \, \tau)+ \kappa \, \tau +\frac{\omega^2}{\Omega} +\partial_\phi \left( \frac{\omega}{\Omega}\right)- \frac{\Omega}{\ell^2}\right) \, .
\end{align}
\end{subequations}

\subsection{Solution phase space, a preliminary discussion}\label{sec:soln-space}

The family of geometries in \eqref{line-element-001} are the most general ones with $r=0$ as the null surface and  in the leading order expansion in $r$ they are specified by seven unknown functions over the null cylinder spanned by $v,\phi$. A special class of these geometries are solutions to  equations of motion \eqref{TMG-eom}.  As will become more explicit and apparent in the coming sections, there are three independent field equations which may also be imposed perturbatively around $r=0$. These lead to three relations among these functions. A generic geometry in the ``solution space'' is hence specified by four unknown function of $v,\phi$ at the $r=0$ null boundary. Higher order terms in the expansion of the metric in powers of $r$ will be fixed by equations of motion at higher orders. As we will see, however, only the lowest order terms given in \eqref{r-expansion-3d'} contribute to the charge analysis around the null boundary ${\cal N}$. 

The metric expansion \eqref{line-element-001} and \eqref{r-expansion-3d'} of course do not depend on the theory and similar ansatz was also used in \cite{Adami:2020ugu} to analyze Einstein-$\Lambda$ theory. In that case the $V_2$ coefficient was also restricted by the equations of motion and the solution space had only three independent functions in it. In the TMG case, however, 
dealing with third order field equations, we have one less constraint at leading order and hence the solution space has four functions in it. The difference between the two cases is due to the existence of a chiral graviton mode in TMG.

The above discussions may be put in a different, and a bit more systematic wording. In 3$d$, metric has  six components. Three of them can be fixed by diffeomorphisms and hence there remains three independent  functions over the spacetime; $h, U, V$ in the metric \eqref{line-element-001}. Equations of motion describing TMG are third order partial differential equations and solutions of this theory will be uniquely determined when nine co-dimension one functions, e.g. functions of $v,\phi$, are specified on the boundary, say $r=0$. On the other hand, canonical analysis of TMG \cite{Buchbinder:1992pe} implies that, contrary to Einstein gravity, there are four second-class constraints in addition to three first-class constraints. Therefore we expect that $5\, (= 3 +\frac{4}{2})$ of co-dimension one functions can be determined in terms of the other four co-dimension one functions via constraint equations. Hence, solution space near null surface  $r=0$ can be uniquely determined through four functions of $v,\phi$. As we will show in the coming sections, this solution space admits a Poisson bracket structure and is hence a solution phase space.

%%%%%%%%%%%%%%%%%%%%%%%%%%%%%%%%%%%%%%%%%%%%%%%%%%%
\subsection{Null boundary symmetries}\label{NBS algebra-sec}
%%%%%%%%%%%%%%%%%%%%%%%%%%%%%%%%%%%%%%%%%%%%%%%%%%%
Consider the vector field $\xi$ whose components are
\begin{equation}\label{xi-sym-gen}
\begin{split}
\xi^v &= T\\
\xi^r &= r (\partial_v T- W) + \frac{r^2 \left( \Omega^2 U_1 + \partial_\phi \eta\right) \partial_\phi T}{ 2\Omega^2}  
%+\frac{1}{3}r^3 \partial_{\phi}T\left(U_2 - \frac{R_1 \partial_\phi \eta}{2 \Omega^4}\right)
+\mathcal{O}(r^3)\\
\xi^\phi & =  Y - \frac{r\eta \partial_\phi T}{\Omega^2}+ \frac{r^2\eta R_1\partial_\phi T}{2 \Omega^4}
%+\frac{r^3 \eta}{3\Omega^4}\left(h_{2}-\frac{h_{1}^{2}}{\Omega^2}\right)\partial_{\phi}T 
+\mathcal{O}(r^3)
\end{split}
\end{equation}
where $T=T(v,\phi), W=W(v,\phi), Y=Y(v,\phi)$ are three arbitrary functions which are $2\pi$ periodic in $\phi$.  While transforming the fields $V, h, U, \eta$, the vector field \eqref{xi-sym-gen} preserves the $r=0$ null surface, i.e. $\delta_\xi g^{rr}|_{r=0}=0$.  
Explicitly, the fields transform as
\begin{subequations}\label{general-field-variations}
    \begin{align}
        	\delta_\xi \eta =& \frac{2 \, \eta}{\chi} \left[ \partial_v \hat{T}- \partial_\phi (\mathcal{U} \hat{T})\right] - \hat{W} \eta + \hat{Y} \partial_\phi \eta {+}\frac{2 \eta {\Omega} \hat{T}}{\chi^2} \, \mathcal{T}_{ll} \\
        	\delta_\xi \mathcal{U} =&  \partial_v \hat{Y} + \hat{Y} \partial_\phi \mathcal{U}- \mathcal{U} \, \partial_\phi \hat{Y} \, ,\\
    \delta_\xi \Omega =&   \hat{T} + \partial_\phi ( \hat{Y} \Omega ) \\
        	 \delta_\xi \mathcal{P} =&- \hat{W}- 2 \partial_\phi \hat{Y} +\hat{Y}\partial_\phi \mathcal{P} {+}\frac{2 {\Omega}\hat{T}\, \mathcal{T}_{ll}}{\chi^2}\\
        	 \delta_\xi \chi= &  \partial_v \hat{T}- \partial_\phi (\mathcal{U} \hat{T})+\partial_\phi (\hat{Y} \chi)\\
        	 \delta_\xi \Upsilon =&  - \hat{T} \partial_\phi \mathcal{P} - 2 \partial_\phi \hat{T}+\hat{Y} \partial_\phi \Upsilon +2 \Upsilon \partial_\phi \hat{Y}+ \Omega \partial_\phi \hat{W} + \frac{2 \Omega \hat{T}}{\chi} \mathcal{T}_{l\phi}  \\
\delta_\xi \tau= & -\frac{\hat{T}}{\chi} \left[\frac{\omega^2}{\Omega} +\partial_\phi \left( \frac{\omega}{\Omega}\right)- \frac{\Omega}{\ell^2}\right] +\partial_\phi (\hat{Y} \tau)-\frac{\tau [\partial_v \hat{T} -\partial_\phi (\mathcal{U} \hat{T})]}{\chi} \nonumber \\
			&-\frac{2 \omega}{\Omega}\partial_\phi \big( \frac{\hat{T}}{\chi}\big)-\partial_\phi \left[ \frac{1}{\Omega} \partial_\phi \big( \frac{\hat{T}}{\chi}\big)\right]{-} \frac{{\Omega}\tau \hat{T}}{\chi^2} \, \mathcal{T}_{ll}{-}\frac{\hat{T}}{{2 \Omega}\chi} \mathcal{T}_{\phi\phi} \\%\\
	    	{\delta_\xi \kappa=}& {\hat{Y}\partial_\phi \kappa+ \partial_v \left( \partial_v T - \mathcal{U} \partial_\phi T+ \kappa T\right) -\mathcal{U}\partial_\phi \left( \partial_v T - \mathcal{U} \partial_\phi T+ \kappa T\right)}
    \end{align}
\end{subequations}
where  hatted-generators are defined as
\begin{equation}\label{hatslicing}
    \hat{W} := W + 2\left( \kappa +\frac{\mathcal{U} \partial_\phi \eta}{2\eta}-\frac{\partial_v \eta}{2 \eta}\right) T- \mathcal{U} \partial_\phi T\, , \qquad \hat{Y} := Y+ \mathcal{U} T \, , \qquad \hat{T}: = T \chi \, .
\end{equation}

\paragraph{Null Boundary Symmetry algebra.} $\xi^\mu$ are ``null boundary Killing vectors'' or null boundary symmetries, i.e. vector field keeping the form of the metric \eqref{line-element-001}. Recalling that equations of motion \eqref{TMG-eom} are covariant, $\xi^\mu$ \eqref{xi-sym-gen} hence rotate us in the solution space.
Therefore, $\xi^\mu$ are symmetries of our setting. By definition, hence, one would expect that the null boundary symmetries should form an algebra. The  symmetry generators are ``field dependent'', i.e. components of $\xi^\mu$ depend on the functions specifying the metric.\footnote{The lowest order terms in $\xi^\mu$ \eqref{xi-sym-gen} are field independent, field dependence appears in higher $r$ orders.} Therefore, when computing the Lie-bracket of the symmetry generators we need to adjust for this field dependence to close onto an algebra \cite{Barnich:2011mi, Compere:2015knw}.  Using the adjusted bracket we find the algebra of null boundary symmetries generating vector fields
\begin{equation}\label{3d-NBS-KV-algebra}
    [\xi(  T_1, W_1, Y_1), \xi( T_2,  W_2, Y_2)]_{_{\text{adj. bracket}}}=\xi(  T_{12}, W_{12}, Y_{12})
\end{equation}
where 
\begin{subequations}\label{W12-T12-Y12}
\begin{align}
     T_{12} &=T_1 \partial_v T_2 - T_2 \partial_v T_1 +Y_1 \partial_\phi T_2 - Y_2 \partial_\phi T_1\\
      W_{12} &=   T_1 \partial_v W_2 - T_2 \partial_v W_1 +Y_1\partial_\phi  W_2- Y_2\partial_\phi  W_1+ \partial_v Y_1 \partial_\phi T_2 -\partial_v Y_2 \partial_\phi T_1 
    \\
 Y_{12} &= Y_1\partial_\phi  Y_2- Y_2\partial_\phi  Y_1+T_1 \partial_v Y_2 - T_2 \partial_v Y_1 \, .
\end{align}
\end{subequations}
This is a Diff$(C_2)\ \oplus$ Weyl$(C_2)$ algebra, where $C_2$ is the cylinder spanned by $v,\phi$; $T, Y$ generate diffeomorphisms on this cylinder and $W$ generates a Weyl scaling along the $r$ direction. This algebra is exactly the same as the one we had in \cite{Adami:2020ugu}. This is of course expected as the above analysis is independent of the theory and only relies on the fact that we are expanding a $3d$ metric around a null surface at $r=0$.

%%%%%%%%%%%%%%%%%%%%%%%%%%%%%%%%%%%%%%%%%%%%%%%%%%%%%%
\section{Vanishing Cotton tensor (VCT) solution phase space and charges }\label{sec:VCT}
%%%%%%%%%%%%%%%%%%%%%%%%%%%%%%%%%%%%%%%%%%%%%%%%%%%%%%%
As discussed all solutions to Einstein-$\Lambda$ theory are also solutions to TMG. This class of solutions have vanishing Cotton tensor and we hence dub them as VCT solutions which satisfy,
\begin{equation}\label{VCT-EoM-full}
 \mathcal{T}_{\mu \nu}=R_{\mu \nu}+ \frac{2}{\ell^2} g_{\mu \nu}=0
 \, ,\qquad C_{\mu\nu}=0  \, .
 \end{equation}
The equations of motion may be decomposed in terms of Raychaudhuri equation $\mathcal{T}_{ll}=0$,  Damour equation $\mathcal{T}_{l \phi}=0$ and $\mathcal{T}_{\phi \phi}=0$. At the zeroth order in $r$, they respectively lead to
\begin{subequations}\label{VCT-EoM}
\begin{align}
& \partial_v \chi - \partial_\phi (\mathcal{U} \chi)-\kappa\ \chi =0\,\label{Raychaudhuri-equation}\\
&   \partial_v \omega - \partial_\phi (\mathcal{U} \omega)+ \frac{\chi \, \omega}{\Omega} - \partial_\phi \kappa =0 \, ,\label{Damour-equation}\\
 &  \partial_v \tau - \partial_\phi (\mathcal{U} \, \tau)+ \kappa \, \tau +\frac{\omega^2}{\Omega} +\partial_\phi \left( \frac{\omega}{\Omega}\right)- \frac{\Omega}{\ell^2}=0 \, .\label{Tmm-equation}
\end{align}
\end{subequations}

The equations of motion \eqref{VCT-EoM-full} restrict the family of near null boundary solutions to exactly those discussed in section 3 of \cite{Adami:2020ugu}, which considered the maximal phase space for $3d$ Einstein-$\Lambda$ around a null surface. Explicitly, the metric \eqref{line-element-001} at first order in $r$ is specified by six functions, $\eta,\Omega,\mathcal{U}$ and $V_1, R_1, U_1$. The latter three are encoded in $\kappa,\tau, \omega$. One may solve for $\kappa,\tau, \mathcal{U}$ in terms of $\eta,\Omega, \omega$ using \eqref{VCT-EoM}. The ${\cal O}(r^2)$ terms will then be determined through the lower order functions using equations of motion \eqref{VCT-EoM-full}, but as will be discussed below, do not contribute to the surface charges at $r=0$.
Therefore, the solution space is specified by three functions over the $r=0$ null surface spanned by $v,\phi$.

\paragraph{Surface charges at the null boundary.} The null boundary Killing vectors \eqref{NBS algebra-sec} generate symmetries over the VCT solution space which is parametrised  by three functions. Since there are three functions in $\xi^\mu$, there are {at most} three associated charges  which are functions over the solution space.  Variation of the Lee-Wald part of charge densities over the VCT solution space for TMG may be computed evaluating  \eqref{TMG-surface-charge}. Straightforward, but tedious computation yields 
\begin{align}\label{charge-VCT case}
   % \begin{split}
  16 \pi G \, \mathcal{Q}^{v r}_{\text{\tiny TMG}}\big|_{r=0} &= 16 \pi G\, \mathcal{Q}^{v r}_{\text{\tiny GR}}\Big|_{r=0} +\frac1\mu\left( \delta \Gamma^{\alpha}_{\phi \beta} \nabla_\alpha \xi^\beta+\frac{1}{\ell^2}h_{\phi \alpha}\xi^{\alpha} \right)\Big|_{r=0} \nonumber \\
    &=\hat{W} \, \delta \left( \Omega + \frac{1}{\mu}\omega - \frac{1}{\mu} \,\frac{\partial_\phi \eta}{\eta} \right) + \hat{Y} \, \delta \left( \Upsilon + \frac{1}{\mu}\varpi \right)  +\hat{T} \delta \mathcal{P} \nonumber\\
    & - \frac{1}{\mu}\partial_v \hat{Y} \, \delta (\tau \,\Omega) + \frac{1}{\mu {\chi}}\left[ -\tau \, \Omega \, \delta \mathcal{U} + \delta \left( \frac{\partial_\phi \eta}{\eta}\right)\right] \left(\partial_v \hat{T}  {+} {\hat T} \, {\mathcal{{T}}_{ll}} {\frac{\Omega}{\chi}}\right)\\
    & +\frac{\hat{T}}{\mu\chi} \biggl\{ \partial_\phi \left[ \frac{\chi \delta \mathcal{P}}{\Omega}- \frac{\partial_\phi (\delta \mathcal{U}\, \Omega)}{\Omega}+2 \delta \mathcal{U}\,\omega \right]+ \chi \,\mathcal{U} \,\partial_\phi \left[\frac{1}{\chi}\left( -\tau \, \Omega \, \delta \mathcal{U} + \delta \left( \frac{\partial_\phi \eta}{\eta}\right) \right)\right] \nonumber\\
    &+\delta \mathcal{U} \left[ \chi \, \tau  - \omega^2- \Omega \partial_\phi \left(\frac{\omega }{\Omega} \right)+ \frac{\Omega^2}{\ell^2}\right]- \frac{\omega \chi \delta \mathcal{P}}{\Omega}+\frac{\chi }{\Omega} \, \partial_\phi \left( \frac{\delta \Omega}{\Omega}\right)  \biggr\}\,.\nonumber
   % \end{split}
\end{align}
The ${\cal T}_{ll}$ term  vanishes on-shell. The above has been written  using the slicing \eqref{hatslicing}. This is the slicing used in \cite{Adami:2020ugu} to render the charges integrable. Due to the presence of last three lines in \eqref{charge-VCT case}, the TMG charges are not integrable in this slicing, nevertheless, we show next that  $Y$-ambiguity can be employed to remedy this.

%%%%%%%%%%%%%%%%%%%%%%%%%%%%%%%%%%%%%%%%%%%%%%%%%%%%%%%%
\newcommand{\hhat}[1]{\hat{\hat{#1}}}
%%%%%%%%%%%%%%%%%%%%%%%%%%%%%%%%%%%%%%%%%%%%%%%%%%%%%%%%
\subsection{Integrability: Fixing the \texorpdfstring{$Y$-ambiguity}{Ya}}
%%%%%%%%%%%%%%%%%%%%%%%%%%%%%%%%%%%%%%%%%%%%%%%%%%%%%%%%
As discussed,   VCT solutions do not have a local (propagating) degree of freedom. Therefore,  there should be a slicing of the solution phase space where the charges become integrable. In the  $\mu \rightarrow \infty$ limit, the  hat-slicing \eqref{hatslicing} works perfectly well and renders the charge integrable, recovering the results of  \cite{Adami:2020ugu}. For finite $\mu$, however, the charges are not integrable in this slicing. We tried many different slicings (reparametrisations on the solution phase space) but the charges remained non-integrable. Nevertheless, this issue can be resolved recalling  the inherent $Y$-ambiguity of the Iyer-Wald procedure, \emph{cf.} \eqref{IWamb}. Remarkably, there exists a $Y$-term which makes  the charges integrable.  Integrability can be used {as a criterion} to partially fix this $Y$-ambiguity.\footnote{We note that  requirement of integrability concerns the finite part of the charges and is different from the prescriptions used in the (holographic) renormalisation procedure to eliminate the divergent part of the charges \cite{Ruzziconi:2020wrb}.  } Besides the addition of the $Y$-term, to arrive at the final conveniently written result we need to make a couple of more changes of slicing, as we outline below.

We now detail the computations.  First we note that the charge density \eqref{charge-VCT case} can be suggestively written as
\begin{equation}\label{appchargedelta}
\begin{comment}
\begin{split}
    &\slashed{\delta} Q_{\text{\tiny LW}}(\xi) {\approx}   \frac{1}{16 \pi G} \int_0^{2 \pi} \d \phi \biggl\{ \hat{Y} \, \delta \tilde{\Upsilon}+\delta_\xi \left(\tilde{\Omega} +\frac{1}{2\mu} \, \partial_\phi \mathcal{P} \right) \, \delta \mathcal{P}-\delta \tilde{\Omega} \, \delta_\xi \mathcal{P}
     \cr & +\frac{1}{\mu} \biggl[ \delta \mathcal{U}\, \delta_\xi (\tau \, \Omega) -\delta_\xi \mathcal{U}\, \delta (\tau \, \Omega)  + \frac{1}{2 \Omega^2} \left( \delta_\xi \Omega \,\partial_\phi \delta \Omega - \delta \Omega \,\partial_\phi \delta_\xi \Omega\right)
    % \\ & \qquad  
    +   \frac{1}{4\eta^2} \left( \delta_\xi \eta \,\partial_\phi \delta \eta - \delta \eta \,\partial_\phi \delta_\xi \eta\right)\biggr]\biggr\},
    \end{split}
\end{equation}
where ${\approx}$ denotes   on-shell  equality in the VCT case.
\cnote{Maybe one would like to write the expression off-shell (note that in the charge formula itself, one has used $R=-6/\ell^2$)? 
\begin{equation}\label{appchargedeltaoffshell}
\end{comment}
\begin{split}
    &\slashed{\delta} Q_{\text{\tiny LW}}(\xi) =   \frac{1}{16 \pi G} \int_0^{2 \pi} \d \phi \biggl\{ \hat{Y} \, \delta \tilde{\Upsilon}+\delta_\xi \left(\hat{\Omega} +\frac{1}{2\mu} \, \partial_\phi \mathcal{P} \right) \, \delta \mathcal{P}-\delta \hat{\Omega} \, \delta_\xi \mathcal{P}
     \cr 
     & +\frac{1}{\mu} \biggl[ \delta \mathcal{U}\, \delta_\xi (\tau \, \Omega) -\delta_\xi \mathcal{U}\, \delta (\tau \, \Omega)  + \frac{1}{2 \Omega^2} \left( \delta_\xi \Omega \,\partial_\phi \delta \Omega - \delta \Omega \,\partial_\phi \delta_\xi \Omega\right)
    % \\ & \qquad  
    +   \frac{1}{4\eta^2} \left( \delta_\xi \eta \,\partial_\phi \delta \eta - \delta \eta \,\partial_\phi \delta_\xi \eta\right)\biggr]\biggr\}\\
    & +\frac{\hat T}{\chi^2}\mathcal T_{ll} \delta\left(\Omega^2\right) +\frac{1}{\mu} \frac{\hat T}{\chi}\biggl[\frac12\mathcal T_{\phi\phi}\delta \mathcal U -\mathcal T_{l\phi} \delta \mathcal P+\mathcal T_{ll}\frac{\Omega}{\chi}\delta\left( \frac{\Upsilon}{\Omega}+\partial_\phi \mathcal P\right) \biggr]\biggr\}
    \end{split}
\end{equation}
%}
where
\begin{align}\label{hat-Omega-tilde-Upsilon-def}\nonumber
  %   \mathcal{P}:= &\ln{\left( \frac{\eta}{\chi^2}\right)}\, , \\ \qquad
     \hat{\Omega} &:= \Omega +\frac{1}{\mu} \, \frac{\Upsilon}{{2}\Omega}\, ,\\
     \tilde{\Upsilon}&:=\Upsilon +2 \partial_\phi \hat{\Omega} +\hat{\Omega} \partial_\phi \mathcal{P}+\frac{1}{\mu}\biggl\{-2 \, \tau\,\chi + \frac{\Omega^2}{\ell^2}+ \left(  \frac{\Upsilon}{2\Omega}\right)^2 + \frac{1}{4}(\partial_\phi \mathcal{P})^2+\partial_\phi^{{2}} \mathcal{P}\biggr\}\,.
     %\\
   %  \varpi:= &-2 \, \tau\,\chi - \tau \, \Omega \, \partial_\phi \mathcal{U}+ \frac{\Omega^2}{\ell^2}+ \left(  \frac{\Upsilon}{2\Omega}\right)^2 + \left(  \frac{\partial_\phi \eta}{2\eta}\right)^2 + 2\left(  \frac{\partial_\phi \Omega}{2\Omega}\right)^2 -\partial_\phi \left(\tau \,\Omega \, \mathcal{U} + \frac{\partial_\phi \Omega}{\Omega} \right) \, .
 \end{align}
From the expression \eqref{appchargedelta}, we see that the first line is integrable if we make the following change of slicing 
\begin{equation}%\label{tilde-basis-1}
\hhat{W}\approx -\delta_\xi \mathcal{P},\qquad \hhat{T}\approx \delta_\xi (\hat{\Omega}+\frac{1}{2\mu} \, \partial_\phi \mathcal{P}  ),\qquad \hhat{Y}:= \hat Y, \qquad \delta \hhat{W}=\delta \hhat{T}=\delta \hhat{Y}=0, \nonumber
\end{equation}
where $\approx$ denotes on-shell equality. The double-hat symmetry generators $\hhat{W},\hhat{T}, \hhat{Y}$ are related to the original symmetry generators $W,T,Y$ through \eqref{hatslicing} where on-shell we drop terms proportional to ${\cal T}_{\mu\nu}$. In the double hat-slicing,  
the second  line of \eqref{appchargedelta} still remains  non-integrable and the last line  vanishes on-shell for the VCT case.

The non-integrable part comes from the CS contribution to the charge density \eqref{CS-contribution}. Its form suggests that a $Y$-term of the following form \begin{equation}\label{YtermB}
    Y^{\mu \nu}[\delta g ; g]= \frac{\sqrt{-g}}{8 \pi G \mu} \, \epsilon^{\mu \nu \lambda} B_\lambda [\delta g ; g]
\end{equation}
with $B_\lambda$ depending on the (variations) of  metric and  Christoffel symbols. Straightforward but lengthy algebra reveals that 
\begin{equation}\label{B}
    B_\lambda [\delta g ; g] = -\frac{1}{8} \,\Gamma^\alpha_{\lambda \beta} \delta g^\beta_\alpha +\frac{1}{2}\, n_\alpha l^\beta \delta \Gamma^{\alpha}_{\lambda \beta} {- \frac{\delta \Omega}{2 \Omega} \partial_\lambda \mathcal{P}} \,,
    % B_\lambda [\delta g ; g] = -\frac{1}{4} \Gamma^\alpha_{\lambda \beta} h^\beta_\alpha + n_\alpha l^\beta \delta \Gamma^{\alpha}_{\lambda \beta}  \, ,
\end{equation}
does the job.\footnote{Note that integrability requirement does not completely fix $Y$-ambiguity; there are other $Y$-terms which make the charges integrable. In essence the $Y$-term \eqref{B} is a minimal choice. }
 The $B_\phi$ component,
\begin{equation}
   2 B_\phi = -\frac{\delta \Omega \partial_\phi \Omega}{2 \Omega^2}-\frac{\delta \eta \partial_\phi \eta}{4 \eta^2}+ \Omega \tau \, \delta \mathcal{U} - \delta \omega { -\frac{\delta \Omega}{\Omega} \partial_\phi \mathcal{P}} +\mathcal{O}(r)\,,
\end{equation}
%\end{subequations}
 contributes to the charge and removes the second line in \eqref{appchargedelta}. Adding $Y$-term \eqref{YtermB} to the surface charge, we have
\begin{equation}
\slashed{\delta} Q(\xi) \approx \frac{1}{16 \pi G} \int_0^{2 \pi} \d \phi \left[\hat{Y} \, \delta \tilde{\Upsilon}  +\delta_\xi \left({\hat \Omega + \frac1\mu \frac{\partial_\phi \Omega}{\Omega} }+\frac{1}{2\mu} \, \partial_\phi \mathcal{P} \right)  \delta \mathcal{P} -\delta_\xi \mathcal{P}\,\delta \left({\hat\Omega + \frac1\mu \frac{\partial_\phi \Omega}{\Omega} }\right) \right].
\end{equation}
Upon a further change of slicing 
$\delta_\xi ({\hat \Omega + \frac1\mu \frac{\partial_\phi \Omega}{\Omega} }+\frac{1}{2\mu} \, \partial_\phi \mathcal{P}  ) \approx \tilde{T}$, the charges become integrable.  Explicitly, we define 
\begin{subequations}\label{vct-int-slicing}
    \begin{align}
        \tilde{W} := & \, \hat{W}+ 2 \partial_\phi \hat{Y} -\hat{Y}\partial_\phi \mathcal{P},\qquad \tilde{Y}:=\hat{Y}, \\
        \tilde{T}  := & \,\frac{\hat{T}}{\Omega} \left(\Omega -\frac{1}{2\mu} \frac{\hat{\Upsilon}}{\Omega} \right) + \partial_\phi \left[ \hat{Y}\left(\Omega +\frac{1}{2\mu} \frac{\hat{\Upsilon}}{\Omega} \right)\right]
    \end{align}
\end{subequations}
where  $\hat{\Upsilon}:= \Upsilon+2 \partial_\phi \Omega + \Omega \partial_\phi \mathcal{P}$. 
In the tilde-slicing surface charges are integrable:
\begin{equation}\label{Integrable-Charge}
    {\delta} Q(\xi) \approx \frac{1}{16 \pi G} \int_0^{2 \pi} \d \phi \left(  \tilde{Y} \, \delta \tilde{\Upsilon}+\tilde{T} \,  \delta \mathcal{P}+\tilde{W}\, \delta \tilde{\Omega} \right),
\end{equation}
where 
\begin{equation}
         \tilde{\Omega} :=%\Omega +\frac{1}{2\mu} \frac{\hat{\Upsilon}}{\Omega} -\frac{1}{2\mu} \, \partial_\phi \mathcal{P} \tcr
         {\hat \Omega + \frac1\mu \frac{\partial_\phi \Omega}{\Omega} }\,.
\end{equation}
The transformation laws are
\begin{equation}\label{tilde-basis-charge-variation-1}
    \delta_\xi \tilde{\Omega} \approx \,\tilde{T}+ \frac{1}{2 \mu} \partial_\phi \tilde{W}\, ,  \qquad \delta_\xi \mathcal{P} \approx \, -\tilde{W} \, ,
\end{equation}
\begin{equation}\label{tilde-basis-charge-variation-2}
    \delta_\xi \tilde{\Upsilon} \approx \, \tilde{Y} \partial_\phi \tilde{\Upsilon} +2 \tilde{\Upsilon} \partial_\phi \tilde{Y} -\frac{2}{\mu} \, \partial_\phi^3 \tilde{Y}\, .
\end{equation}
Also, for later use we record that equations of motion imply
\begin{equation}
   \partial_v \tilde{\Upsilon} \,\approx\, \mathcal{U} \partial_\phi \tilde{\Upsilon} +2 \tilde{\Upsilon} \partial_\phi \mathcal{U} -\frac{2}{\mu} \, \partial_\phi^3 \mathcal{U}.
\end{equation}

%%%%%%%%%%%%%%%%%%%%%%%%%%%%%%%%%%%%%%%%%%%%%%%%%%%%%%%%
\subsection{Algebra of symmetry generators in the integrable slicing}
%%%%%%%%%%%%%%%%%%%%%%%%%%%%%%%%%%%%%%%%%%%%%%%%%%%%%%%%
The algebra of boundary symmetry generators in $W$, $Y$ and $T$ slicing is given in \eqref{3d-NBS-KV-algebra} and \eqref{W12-T12-Y12}. The algebra in the hatted or tilde -slicing where $\hat{W}$, $\hat{Y}, \hat{T}$ or $\tilde{W}$, $\tilde{Y}, \tilde{T}$ are treated as field independent functions, is different. To account for the field dependence of the transformation from the original slicing to these new slicing one need to use the adjusted Lie bracket \cite{Barnich:2011mi, Compere:2015knw}. 

\paragraph{Null boundary symmetry algebra in the hatted-slicing:}
\begin{equation}\label{NHKV-algebra-3d}
	[\xi( \hat T_1, \hat W_1, \hat Y_1), \xi( \hat T_2, \hat W_2, \hat Y_2)]_{_{\text{adj. bracket}}}=\xi( \hat T_{12}, \hat W_{12}, \hat Y_{12}).
\end{equation}
where
	\begin{subequations}\label{NHKV-algebra-3d-01}
		\begin{align}
		& \hat T_{12}=\partial_\phi (\hat Y_1 \hat T_2-\hat Y_2 \hat T_1),\\
		& \hat W_{12}= \hat{Y}_1\partial_\phi \hat W_2 -\hat{Y}_2\partial_\phi \hat W_1,
		\\
		& \hat Y_{12}=\hat Y_1\partial_\phi \hat Y_2-\hat Y_2\partial_\phi \hat Y_1.
		\end{align}
	\end{subequations}
The algebra is exactly the same as what was obtained in section 3 of \cite{Adami:2020ugu} for $s=0$. 

\paragraph{Null boundary symmetry algebra in the tilde-slicing:}
\begin{equation}\label{NHKV-CVT-tilde-basis-1}
	[\xi( \tilde T_1, \tilde W_1, \tilde Y_1), \xi( \tilde T_2, \tilde W_2, \tilde Y_2)]_{_{\text{adj. bracket}}}=\xi( \tilde T_{12}, \tilde W_{12}, \tilde Y_{12}).
\end{equation}
where
	\begin{equation}\label{NHKV-CVT-tilde-basis-2}
 \tilde W_{12}=0, \hspace{1 cm}\tilde T_{12}= 0,  \hspace{1 cm} \tilde Y_{12}=\tilde Y_1\partial_\phi \tilde Y_2-\tilde Y_2\partial_\phi \tilde Y_1.
	\end{equation}
This  is a  $ {\cal A}_2\ \oplus$ Witt algebra; the  Witt part is generated by $\tilde Y$, the ${\cal A}_2$  denotes local $u(1)\oplus u(1)$ algebra on the null cylinder and is generated by $\tilde W, \tilde T$.

%%%%%%%%%%%%%%%%%%%%%%%%%%%%%%%%%%%%%%%%%%%%%%%%%%%%%%%%
\subsection{Charge algebra in the integrable slicing}\label{sec:charge-algebra-VCT}
%%%%%%%%%%%%%%%%%%%%%%%%%%%%%%%%%%%%%%%%%%%%%%%%%%%%%%%%

As discussed and may be explicitly seen in \eqref{Integrable-Charge}, the tilde-slicing leads to integrable charges ${\cal P}, \tilde\Omega, \tilde\Upsilon$, respectively associated with $\tilde T, \tilde W, \tilde Y$ generators. According to the fundamental theorem of the covariant phase space method \cite{Lee:1990nz, Iyer:1994ys}, the algebra of charges is the same as the algebra of symmetry generators \eqref{NHKV-CVT-tilde-basis-1}, \eqref{NHKV-CVT-tilde-basis-2}, possibly up to central terms. Since we have the explicit form of the charges and their transformations \eqref{tilde-basis-charge-variation-1}, \eqref{tilde-basis-charge-variation-2} we can compute the charge algebra and explicitly confirm  the stated theorem:
\begin{equation}\label{charge-algebra-CVT-01}
{\delta_{\xi_2} Q_{\xi_1}=}    \{Q_{\xi_1},Q_{\xi_2}\}=Q_{\xi_{12}}+K_{\xi_1,\xi_2}
\end{equation}
where the central extension term $ K_{\xi_1,\xi_2}$ is
\begin{equation}\label{central-extension-term}
    K_{\xi_1,\xi_2}=\frac{1}{16\pi G}\int_{0}^{2\pi}d\phi\left[\left(\tilde{W}_{1}\tilde{T}_2
    %-\tilde{W}_2\tilde{T}_{1}\right)
    -\frac{1}{\mu}\tilde{Y}_{1}\partial_{\phi}^3 \tilde{Y}_{2}
    %-\tilde{Y}_{2}\partial_{\phi}^3\tilde{Y}_{1}\right)
    +\frac{1}{4\mu}\tilde{W}_{1}\partial_{\phi}\tilde{W}_{2}\right)-\left(1\leftrightarrow 2\right)%\tilde{W}_{2}\partial_{\phi}\tilde{W}_{1}\right)
    \right].
\end{equation} 
The charge algebra in terms of $\tilde{\Omega}$, $\tilde\Upsilon$ and $\mathcal{P}$ is
\begin{subequations}\label{charge-algebra-IB-02}
		\begin{align}
		&\{\mathcal{P}(v,\phi),\mathcal{P}(v,\phi')\}=0,\qquad
		\{\tilde{\Omega}(v,\phi),\mathcal{P}(v,\phi')\}=16\pi G\delta(\phi-\phi'),\\
		&\{\tilde{\Omega}(v,\phi),\tilde{\Omega}(v,\phi')\}=\frac{8\pi G}{\mu}\partial_{\phi}\delta(\phi-\phi'),\\
		& \{\tilde\Upsilon(v,\phi),\tilde\Omega(v,\phi')\}=0,\qquad
		\{\tilde\Upsilon(v,\phi),\mathcal{P}(v,\phi')\}=0,\\
		&\{\tilde{\Upsilon}(v,\phi),\tilde{\Upsilon}(v,\phi')\}=16\pi G\left(\tilde{\Upsilon}(v,\phi')\partial_{\phi}-\tilde{\Upsilon}(v,\phi)\partial_{\phi'}-\frac{2}{\mu}\partial_{\phi}^3\right)\delta(\phi-\phi').
		\end{align}
\end{subequations}	
In the large $\mu$ limit, as expected, we obtain the Heisenberg $\oplus$ Diff($S^1$) algebra \cite{Adami:2020ugu}. There are three central terms in the  charge algebra \eqref{charge-algebra-IB-02}, two of which are proportional to $1/\mu$ and hence appear due to the presence of the CS term. 

The algebra in the the above slicing is not yet in our final convenient form. One of the central terms can still be removed through another change of slicing:
\begin{equation}\label{calS-def}
    {\cal S}:%=\bar\Omega
    =\tilde\Omega+\frac{1}{4\mu}\partial_{\phi}\mathcal{P} { = \Omega +\frac{1}{2\mu\Omega}\left( \, {\Upsilon} + 2\partial_\phi \Omega +\frac12\Omega\partial_{\phi}\mathcal{P}\right)}\,.
\end{equation}
In this slicing the charge expression yields
\begin{equation}\label{VCTintcharge}
    \delta Q_{\xi}=\frac{1}{16\pi G}\int_{0}^{2\pi}\d \phi \, (\tilde{Y}\delta{\cal J}+\tilde{\mathbb{T}}\delta\mathcal{P}+\tilde{W}\delta{\cal S})
\end{equation}
where
\begin{equation}\label{calJ-bbT-def}
{\cal J}:=\tilde{\Upsilon},\qquad \tilde{\mathbb{T}}:=\tilde{T}+\frac{1}{4\mu}\partial_{\phi}\tilde{W},
\end{equation}
and variation of ${\cal S}$ gives the simple form
\begin{equation}
    \delta_{\xi} {\cal S}=\tilde{\mathbb{T}}.
\end{equation}
	
The null boundary symmetries algebra in the new tilde-slicing is
\begin{equation}\label{tilde-basis-adj-bracket}\begin{split}
[\xi(\tilde{\mathbb{T}}_{1}, \tilde W_1, \tilde Y_1), \xi( \tilde{\mathbb{T}}_2, \tilde W_2,\tilde Y_2)]_{_{\text{adj. bracket}}}=\xi( \tilde{\mathbb{T}}_{12}, \tilde W_{12}, \tilde Y_{12}),\\
 \tilde W_{12}=0, \hspace{1 cm} \tilde{\mathbb{T}}_{12}= 0,  \hspace{1 cm} \tilde Y_{12}=\tilde Y_1\partial_\phi \tilde Y_2-\tilde Y_2\partial_\phi \tilde Y_1.
\end{split}
	\end{equation}
In this slicing the algebra becomes a Heisenberg $\oplus$ Virasoro algebra:
\begin{subequations}\label{VCT-algebra-SPJ}
    \begin{align}
      \hspace*{-8mm}  \{{\cal S}(v,\phi),{\cal S}(v,\phi')\} =0,\ 
		\{{\cal S}(v,\phi),\mathcal{P}(v,\phi')\}&=16\pi G\delta(\phi-\phi'), \ 	\{\mathcal{P}(v,\phi),\mathcal{P}(v,\phi')\}=0,\\
			 \{{\cal J}(v,\phi),{\cal S} (v,\phi')\}=0,&\qquad
		\{{\cal J}(v,\phi),\mathcal{P}(v,\phi')\}=0,\\
		\{{\cal J}(v,\phi),{\cal J}(v,\phi')\} =16\pi G\bigg(&{\cal J}(v,\phi')\partial_{\phi}-{\cal J}(v,\phi)\partial_{\phi'}-\frac{2}{\mu}\partial_{\phi}^3\bigg)\delta(\phi-\phi'). \end{align}
\end{subequations}
In terms of Fourier modes of the charges,
\begin{equation}
\boldsymbol{\mathcal{X}}_n:=\frac{1}{16\pi G}\int_0^{2\pi} \d{}\phi\ \mathcal{X}(v,\phi) e^{in\phi},
\end{equation}
after ``quantizing the charges'' upon replacing $\{\cdot, \cdot\}\to -i[\cdot, \cdot]$, we obtain
\begin{subequations}\label{VCT-algebra-SPJ-Fourier}
    \begin{align}
      \hspace*{-8mm}  [\boldsymbol{\mathcal{S}}_n,\boldsymbol{\mathcal{S}}_m] =0,\quad 
		[\boldsymbol{\mathcal{S}}_n,\boldsymbol{\mathcal{P}}_m]&=\frac{{i}}{8G}\delta_{m+n,0}, \quad 	[\boldsymbol{\mathcal{P}}_n,\boldsymbol{\mathcal{P}}_m]=0,\\
			 [\boldsymbol{\mathcal{J}}_n,\boldsymbol{\mathcal{S}}_m]=0,&\qquad
		 [\boldsymbol{\mathcal{J}}_n,\boldsymbol{\mathcal{P}}_m]=0,\\
		 [\boldsymbol{\mathcal{J}}_n,\boldsymbol{\mathcal{J}}_m] =&(n-m) \boldsymbol{\mathcal{J}}_{n+m} +\frac{1}{4\mu G}n^3\delta_{n+m,0}. \label{VCT-algebra-Virasoro}
    \end{align}
\end{subequations}
We note that the Fourier modes $\boldsymbol{\mathcal{S}}_n, \boldsymbol{\mathcal{P}}_n, \boldsymbol{\mathcal{J}}_n$, respectively entropy aspect charge, expansion aspect charge and angular momentum aspect charge, are in general functions of $v$. The zero mode of $\boldsymbol{\mathcal{S}}$ is equal to the Wald entropy \cite{Wald:1993nt, Iyer:1994ys}, $\boldsymbol{\mathcal{P}}_0$ is proportional to the logarithm of the expansion of the null vector field generating the null boundary and $\boldsymbol{\mathcal{J}}_0$ is the angular momentum charge.

\paragraph{On central charges of the charge algebra.} {There are two kinds of central terms in our case, one is is $1/(8G)$ which appears in the Heisenberg part   and is $\mu$ independent. The other comes from the CS term and appears in the Virasoro part. The former is also present in the $3d$ Einstein gravity analysis \cite{Adami:2020ugu}.} The central term in the Virasoro algebra \eqref{VCT-algebra-SPJ-Fourier} is coming from the gravitational (diffeomorphism) anomaly in the presumed dual 2d CFT \cite{Kraus:2005vz, Kraus:2005zm, Solodukhin:2005ah}. To see this, let us first note \eqref{xi-sym-gen} and that Virasoro part is associated with ``super-rotations''. Next, recall the asymptotic symmetry analysis of TMG with  Brown-Henneaux boundary conditions yields two left and right Virasoro algebras with central charges \cite{Hotta:2008yq, Compere:2008cv}, $c_L=\frac{3\ell}{2G}+\frac{3}{2\mu G} ,c_R=\frac{3\ell}{2G}-\frac{3}{2\mu G}$. The difference of the left and right Virasoro algebras generate super-rotations, which is a Virasoro algebra at central charge $c_L-c_R=\frac{3}{\mu G}$, which exactly matches the central charge in \eqref{VCT-algebra-Virasoro}.

\subsection{Non-conservation of charges and generalized conservation equation}
\label{sec:GCCE-4.4}

As discussed integability and conservation of surface charges are known to be closely related to each other. This latter has been noted e.g. in Wald-Zoupas  \cite{Wald:1999vt} and also in the Barnich-Troessaert \cite{Barnich:2011mi} analyses and more recently in \cite{Adami:2020amw}. Here we revisit integrability and conservation in presence of central charges. To this end, we  study ``time evolution'' of the charges. Our charges are functions of light-cone time coordinate $v$, as they are given by integrals over $\phi$ on integrands which are functions of $v,\phi$.  $\partial_v$ is not among the generators of the tilde-slicing in which the charges are integrable.\footnote{$\partial_v$ is a vector corresponding to $T=1, Y=W=0$, for which $ \tilde{\mathbb{T}}^{(v)}{\approx}  \partial_v \mathcal{S},\ \tilde W^{(v)}{\approx} -\partial_v \mathcal{P}, \  \tilde Y^{(v)}=\mathcal{U}. $ 
$\partial_v \mathcal{S}, \partial_v \mathcal{P}$ are related to the other fields and their $\phi$ derivatives using the on-shell conditions. That is, $\partial_v$ is field-dependent in the tilde-slicing. While one cannot define the charge associated with $\partial_v$, the corresponding charge variation is well-defined. \label{v-vs-tilde v-footnote}} Nonetheless, one can locally introduce another appropriate ``time coordinate'' $\tilde v$ more closely related to the tilde-slicing: ${\xi}(\tilde{\mathbb{T}},\tilde W,\tilde Y)={\xi}(1,0,0):=\partial_{\tilde v}$.

The Hamiltonian evolution equation with respect to $\tilde v$ for a generic charge $Q^{\text{I}}(\xi)$ is 
\begin{equation}\label{hamilevoleq}
   \frac{\d{}}{\d {\tilde v}} Q^{\text{I}}(\xi) = \delta_{\partial_{\tilde v}}Q^{\text{I}}_{\xi} 
   +\partial_{\tilde v} Q^{\text{I}}_{\xi} \, ,\qquad \delta_{\partial_{\tilde v}}Q^{\text{I}}_{\xi} :=\{Q^{\text{I}}(\xi), Q^{\text{I}}(\partial_{\tilde{v}})  \}
\end{equation}
The $\partial_{\tilde v} Q^{\text{I}}_{\xi} $ term takes into account the explicit $\tilde v$ dependence which in part comes from  symmetry generator $\xi$.
Using \eqref{charge-algebra-CVT-01} in the tilde-slicing \eqref{VCT-algebra-SPJ} for $\xi_1=\xi (\tilde{\mathbb{T}},\tilde W,\tilde Y)$ arbitrary and $\xi_2=\partial_{\tilde v}$, one has
\begin{equation}\label{GCCE-Integrable}
\frac{\d{}}{\d {\tilde v}} Q^{\text{I}}_{\xi}=\frac1{{16} \pi G} \int_0^{2\pi} \d \phi\ \tilde{W} + \partial_{\tilde v} Q^{\text{I}}_{\xi}.
\end{equation}
Following \cite{Adami:2020amw}, we name this equation as generalized charge conservation equation (GCCE). 
%The last term, we expect to be zero if $\xi$ components have no explicit $\tilde v$ dependence. %
This equation shows that due to the  central term in the Heisenberg part the charge associated with $\xi=\xi(0,\tilde W,0)$ while integrable, is not conserved, whereas the charge associated with $\xi=\xi(\tilde{\mathbb{T}},0,\tilde Y)$ are both integrable and conserved. 

The other central charge in the Virasoro part of the algebra \eqref{VCT-algebra-SPJ-Fourier} does not appear in the GCCE \eqref{GCCE-Integrable}, as this equation captures evolution in $\tilde v$. One can see the effects of this other central charge by studying ``evolution'' in $\tilde \phi$ direction, such that 
$\partial_{\tilde \phi}=\xi(0,0,1)$ in the tilde-slicing. Similarly one could have locally defined another $\tilde r$ coordinate such that  $\tilde r\partial_{\tilde r}=\xi(0,1,0)$ in the tilde-slicing. The ``evolution'' in $\ln\tilde r$ would again involve the central charge in the Heisenberg part. All in all, the central charges yields a non-conservation in the associated charges. This matches with the usual statement that central charges are ``anomalies'' for conserved charges.

%%%%%%%%%%%%%%%%%%%%%%%%%%%%%%%%%%%%%%%%%%%%%%%%%%%%%
\subsection{Non-expanding backgrounds, example of BTZ}\label{sec:non-expanding-case}
%%%%%%%%%%%%%%%%%%%%%%%%%%%%%%%%%%%%%%%%%%%%%%%%%%%%%

So far we discussed the most general solution phase space and charges in the VCT sector. An important subspace arises when the background is non-expanding $\chi=0$, where $\chi$ is defined in \eqref{Theta-chi-def}. Note that this is consistent with the equations of motion \eqref{VCT-EoM}. An important example in this class is when the null surface is a Killing horizon, like the cases in Einstein gravity discussed in \cite{Grumiller:2019fmp}. The phase space of the non-expanding sector cannot be directly  derived from the generic case, as in the vanishing expansion limit, ${\cal P}$ \eqref{calP-def} and  the change of slicing to $\hat T$  \eqref{hatslicing} {become} ill-defined. We need to revisit our charge analysis for such cases keeping $T$ untouched.

While for the background $\chi=0$, one may allow generic perturbations for which $\chi\neq 0$. For $\chi=0$ field variations \eqref{general-field-variations} generated by the null boundary symmetry $\xi$ \eqref{xi-sym-gen}  take the form
\begin{equation}\label{non-expanding-field-variations}\begin{split}
        \delta_\xi \Omega =& \partial_\phi (\hat{Y}\Omega)\\
        \delta_\xi \mathcal{U} =&  \partial_v \hat{Y} + \hat{Y} \partial_\phi \mathcal{U}- \mathcal{U} \, \partial_\phi \hat{Y} \, ,\\
        \delta_\xi \eta  = & - \hat{W} \eta +2 \eta \partial_v T +\hat{Y}\partial_\phi \eta +2 \eta \kappa T -2 \mathcal{U} \eta \partial_\phi T \\
        \delta_\xi \chi =& 0\\
       \delta_\xi \tau = & T \partial_v \tau - \tau \partial_v T + \tau \mathcal{U} \partial_\phi T - T \partial_\phi (\tau \mathcal{U}) + \partial_\phi (\hat{Y}\tau) - \frac{2 \omega \partial_\phi T}{\Omega} - \partial_\phi \left( \frac{\partial_\phi T}{\Omega}\right)\\
       \delta_\xi \Upsilon =&  \hat{Y} \partial_\phi \Upsilon +2 \Upsilon \partial_\phi \hat{Y}+ \Omega \partial_\phi \hat{W} + 2 \Omega {T} \mathcal{T}_{l\phi},\\ \delta_\xi \kappa= & {\hat{Y}\partial_\phi \kappa+ \partial_v \left( \partial_v T - \mathcal{U} \partial_\phi T+ \kappa T\right) -\mathcal{U}\partial_\phi \left( \partial_v T - \mathcal{U} \partial_\phi T+ \kappa T\right)} 
    \end{split}
\end{equation}
where $\hat W, \hat Y$ are defined as in \eqref{hatslicing}. Note that the null boundary symmetry generators cannot take us to a nonzero $\chi$ values. Nonetheless we may still allow $\delta\chi\neq 0$, while computing the charge variations \eqref{TMG-surface-charge}.  One has
\begin{equation}
    \begin{split}
        16 \pi G \, \delta\mathcal{Q}^{v r}_{\text{\tiny TMG}}\big|_{r=0} & \approx\ \hat{W} \, \delta \left( \Omega + \frac{1}{\mu}\frac{\Upsilon}{2 \Omega} \right)+ \hat{Y} \delta\left\{  \Upsilon + \frac{1}{\mu} \left[ -2 \tau 
         \chi + \left( \frac{\Omega^2}{\ell^2}+ \frac{\Upsilon^2}{4 \Omega^2}\right)\right]\right\} \\
        &+ T \delta\left[ -2 \partial_\phi \left( \frac{ \chi}{\Omega}\right)+\frac{2 \omega  \chi}{\Omega}\right]\\
       & +\frac{1}{\mu} \biggl[ \delta \mathcal{U}\, \delta_\xi (\tau \, \Omega) -\delta_\xi \mathcal{U}\, \delta (\tau \, \Omega)  + \frac{1}{2 \Omega^2} \left( \delta_\xi \Omega \,\partial_\phi \delta \Omega - \delta \Omega \,\partial_\phi \delta_\xi \Omega\right) \\ & \qquad  
   +   \frac{1}{4\eta^2} \left( \delta_\xi \eta \,\partial_\phi \delta \eta - \delta \eta \,\partial_\phi \delta_\xi \eta\right)\biggr]\,.
   \end{split}
\end{equation}
The last two lines may be removed by the $Y$-term \eqref{B} without the last term. 
The integrable part of the charge is hence given by
\begin{subequations}\label{Integrable-Charge-non-expanding}
\begin{align}
    Q(\hat{W}) &\approx \frac{1}{16 \pi G} \int_0^{2 \pi} \d \phi\  
    \hat{W}\ \left( \Omega + \frac{1}{\mu}\frac{\Upsilon}{2 \Omega} \right),\\
      Q(\hat{Y}) &\approx 
      %\frac{1}{16 \pi G} \int_0^{2 \pi} \d \phi\ \hat{Y}\ \left\{  \Upsilon + \frac{1}{\mu} \left[ -2 \tau   \chi + \left( \frac{\Omega^2}{\ell^2}+ \frac{\Upsilon^2}{4 \Omega^2}\right)\right]\right\}= 
      \frac{1}{16 \pi G} \int_0^{2 \pi} \d \phi\ \hat{Y}\ \left[  \Upsilon + \frac{1}{\mu} 
         %\left( -2 \tau \chi + 
         \left( \frac{\Omega^2}{\ell^2}+ \frac{\Upsilon^2}{4 \Omega^2}\right)\right], \\
                Q(T) &\approx \frac{1}{16 \pi G} \int_0^{2 \pi} \d \phi\ T\ \left[ -2 \partial_\phi \left( \frac{ \chi}{\Omega}\right)+\frac{2 \omega  \chi}{\Omega}\right] =0\,.
        \end{align}
        \end{subequations}
As we see the charge associated with $T$ is vanishing and is hence a trivial transformation. We may use this ``gauge freedom'' to set $\eta=1$.\footnote{This may be done setting $\delta_\xi\eta=0, \eta=1$ and solving for $T$. See section 6.3 of \cite{Adami:2020amw} for a similar discussion.} The ``non-expanding physical solution phase space'' is then described by the two charges:
\begin{subequations}\label{SJ-non-expanding}
    \begin{align}
\hat{\mathcal{ S}} & :=\Omega + \frac{1}{\mu}\frac{\Upsilon}{2 \Omega}, \\
    \hat{\mathcal{J}} & :=\Upsilon + \frac{1}{\mu} \left( \frac{\Omega^2}{\ell^2}+ \frac{\Upsilon^2}{4 \Omega^2}\right), %- \mu \hat{\mathcal{ S}}^2, \tcr{=\mu {\hat \Omega}^2- (\mu-\frac{1}{\mu\ell^2}) {\Omega^2}}\, ,
   \end{align}
\end{subequations}
with the transformation laws 
\begin{subequations}
    \begin{align}
        \delta_\xi \hat{\mathcal{ S}} \approx\ & \partial_{\phi}(\hat{Y} \hat{\mathcal{ S}})+ \frac{1}{2\mu} \partial_{\phi}\hat{W}\,\\
        \delta_\xi \hat{\mathcal{J}} \approx\ & \hat{Y} \partial_\phi \hat{\mathcal{J}} + 2 \hat{\mathcal{J}} \partial_\phi \hat{Y}+\hat{\mathcal{ S}} \partial_{\phi}\hat{W} \, .
    \end{align}
\end{subequations}
These transformation laws imply the following algebra
\begin{subequations}\label{non-expanding-algebra-1}
		\begin{align}
		&\{\hat{\mathcal{ S}}(v,\phi), \hat{\mathcal{ S}}(v,\phi')\}=\frac{8\pi G}{\mu}\partial_{\phi}\delta(\phi-\phi'),\\
		&\{\hat{\mathcal{J}}(v,\phi), \hat{\mathcal{ S}}(v,\phi')\}= 16\pi G \, \hat{\mathcal{ S}}(v, \phi) \partial_{\phi}\delta(\phi-\phi') ,\\
		&\{\hat{\mathcal{J}}(v,\phi),\hat{\mathcal{J}}(v,\phi')\}=16\pi G\left(\hat{\mathcal{J}}(v,\phi')\partial_{\phi}-\hat{\mathcal{J}}(v,\phi)\partial_{\phi'}\right)\delta(\phi-\phi').
		\end{align}
\end{subequations}
The above algebra is a $U(1)$ Kac-Moody algebra with the $U(1)$ current $\hat{\mathcal{S}}$, at level $1/(16\mu G)$  and the Virasoro part, which is generated by $\hat{\mathcal{J}}$, has vanishing central charge. This is to be contrasted with the algebra of the generic case \eqref{VCT-algebra-SPJ} where the Virasoro part has a central charge equal to $3/(\mu G)$. Absence of the Virasoro central charge in the non-expanding case can be traced to  the fact that this central charge is arising from $\partial^2_\phi {\cal P}$ term in \eqref{hat-Omega-tilde-Upsilon-def} and that ${\cal P}$ vanishes in the non-expanding case. As discussed this central charge is related to the gravitational anomaly of the presumed dual $2d$ CFT. One may then check that for the non-expanding case the diffeomorphism non-invariance of the CS term vanishes for a null boundary with zero expansion $\Theta_{_{l}}$.

It is also interesting to note that upon the redefinition $\hat{\mathcal{J}}\to \hhat{\mathcal{J}}:= \hat{\mathcal{J}}-\mu\hat{\mathcal{ S}}^2$, the algebra takes a simpler form as $\{\hhat{\mathcal{J}}, \hat{\mathcal{S}}\}=0$. However, recalling \eqref{SJ-non-expanding}, $\hhat{\mathcal{J}}=(\frac1{\mu\ell^2}-\mu)\Omega^2$. At the chiral point $\mu\ell=1$ \cite{Li:2008dq}, the algebra degenerates, as $\hat{\mathcal{J}}=\mu\hat{\mathcal{S}}^2$ and one remains with a single independent charge.\footnote{As a side remark we note that at the chiral point $\mu\ell=1$, the expression for $\tilde{\Upsilon}$ in generic case \eqref{hat-Omega-tilde-Upsilon-def} takes a much simpler form  $\mu\tilde{\Upsilon}=\frac14 {\hat{\cal P}}^2+ \partial_\phi {\hat{\cal P}}-2\tau\chi$, where ${\hat{\cal P}}:=2\mu\hat{\Omega} + \partial_\phi \mathcal{P}$. Thus, the algebra at the critical point does not  degenerate in the generic case due to the $-2\tau\chi$ term. \label{footnote12}} 

An example for the non-expanding cases is when the $r=0$ null surface is  Killing horizon of a black hole. These cases were studied in \cite{Grumiller:2019fmp} for Einstein gravity in diverse dimensions. It is readily seen that the ``BMS-like slicing'' with $s=0$ in that work matches the $\mu\to\infty$ limit of \eqref{Integrable-Charge-non-expanding}.\footnote{Note that ${\cal P}$ of  \cite{Grumiller:2019fmp} corresponds to ${\cal S}$ in this work.}

\paragraph{BTZ example.}
As an illustrative example of such backgrounds, we consider the BTZ black hole background \cite{BTZ} which is a solution in the VCT class. Let us choose $r=0$ null surface to  be the outer Killing horizon of the black hole. In our $v,r,\phi$ coordinate system the metric takes the form,
\begin{align}\label{BTZ-metric}
    \d s^2=-\frac{r}{\ell^2}(r+2r_+ )\left(1-\frac{r_-^2}{(r+r_+)^2}\right)\d v^2+2 \d v\d r  +(r+r_+)^2 \left(\d \phi- \frac{r_+\, r_-}{\ell(r+r_+)^2 }\d v\right)^2,
\end{align}
where $r_\pm$ are the horizon radii. In terms of the near $r=0$ expansion \eqref{line-element-001}, one has 
\begin{equation}\begin{split}
& \eta=1, \qquad \Omega=r_+\,,\qquad R_1=2r_+\,,\\ %\quad h_2=1 \,, \, \\
&\mathcal U=-\frac{r_-}{r_+\,\ell}\,,\quad U_1=-\frac{R_1\,\mathcal U}{\Omega^2}=\frac{2r_-}{\ell\,r_+^2}\,, \\
%\quad U_2=-\frac{h_2\,\mathcal U +R_1 \, U_1}{\Omega^2}=-3\frac{r_-}{\ell\,r_+^3}\,, \\ 
& V_1=2\frac{r_+^2-r_-^2}{\ell^2\,r_+}=2\kappa\,, \quad   V_2=\frac{1}{\ell^2}(1+\frac{3r_-^2}{r_+^2}), 
\end{split}\end{equation}
where $\kappa=\kappa_{_{\text{H}}}$ is the surface gravity and $\mathcal U$ is the horizon angular velocity. Thus the quantities 
appearing in the redefinition of  symmetry generators are 
\begin{equation}
\begin{split}
&    \Upsilon=-\frac{2r_+\,r_-}{\ell}\,, %\Gamma=-2\kappa \,, \omega^\pm=-\frac{r_-}{\ell} \,,
    \quad\chi=0\,, \quad  \tau=1\,,\quad \varpi=\frac{r_+^2+r_-^2}{\ell^2}\,, \quad \omega=-\frac{r_-}{\ell}.
   % \\ & \tilde \Omega=r_+ -\frac{r_-}{\mu\ell},\qquad \tilde \Upsilon=-\frac{2r_+ r_-}{\ell}+ \frac{r_+^2+r_-^2}{\ell^2\mu} ,\qquad\omega=-\frac{r_-}{\ell}.
\end{split}
\end{equation}
As a simple check, we note that the field equations \eqref{Raychaudhuri-equation} and \eqref{Damour-equation} are trivially satisfied while \eqref{Tmm-equation} is satisfied as $\kappa=\Omega/\ell^2-\omega^2/\Omega$.

Therefore the background charges \eqref{SJ-non-expanding} are 
\begin{align}
{\cal S}_{\text{\tiny BTZ}}&=2 \pi \, Q(0,\hat{W},0)= \frac{2\pi}{4 G}\left(r_+ -\frac1{\mu\,\ell}r_- \right) \hat W^0(v)\,,\\
{\cal J}_{\text{\tiny BTZ}}&= -Q(0,0,\hat{Y})= \frac{1}{ 4 G}\left(\frac{r_+ r_-}{\ell}- \frac{r_+^2+r_-^2}{2\ell^2\mu}\right)  \hat{Y}^0(v) %J^{\text{\tiny TMG}}-\frac1{4G \mu} r_+ \partial_v \hat Y_0(v) \,.\\
%\Pi_{\text{\tiny BTZ}}&=0,
\end{align}
where $\hat W^0(v), \hat Y^0(v)$ are the Fourier zero modes of the associated symmetry generators. ${\cal S}_{\text{\tiny BTZ}}, {\cal J}_{\text{\tiny BTZ}}$  are respectively proportional to the Iyer-Wald entropy and angular momentum of BTZ black hole in TMG.\footnote{As a side remark, we note that these expressions can be written as ${\cal S}_{\text{\tiny BTZ}}= (S_+ -\frac{ S_-}{\mu\ell}) \hat{W}^0(v)$, ${\cal J}_{\text{\tiny BTZ}}=(J-\frac{M}{\mu\ell})\hat{Y}^0(v)$, where $S_\pm$ are the entropy of the outer and inner horizon and $J,M$ are  angular momentum and mass of a BTZ black hole of horizon radii $r_\pm$ in the usual Einstein AdS$_3$ gravity.}

%%%%%%%%%%%%%%%%%%%%%%%%%%%%%%%%%%%%%%%%%%%%%%%%%%%%%%
\section{ Non-vanishing Cotton Tensor (NVCT) solution phase space and charges}\label{sec:NVCT}
%%%%%%%%%%%%%%%%%%%%%%%%%%%%%%%%%%%%%%%%%%%%%%%%%%%%%%%
After the warm-up, illuminating case of vanishing Cotton tensor solution phase space, we now consider the phase space for generic TMG solutions, when Cotton tensor does not vanish. This case is important and interesting as it allows for propagating massive chiral gravitons and hence non-zero fluxes are expected. 

%%%%%%%%%%%%%%%%%%%%%%%%%%%%%%%%%%%%%%%%%%%%%%%%%%%%%%%%%%%%%%%%%%%
\subsection{Equations of motion}
%%%%%%%%%%%%%%%%%%%%%%%%%%%%%%%%%%%%%%%%%%%%%%%%%%%%%%%%%%%%%%%%%%%

Our starting point is again  metric \eqref{line-element-001} and the conventions sets in section \ref{sec:NNB-metric}. The equations of motion \eqref{extra-condition} ${\cal E}_{\mu\nu}=0$, for NVCT case to lowest order in $r$, give
\begin{equation}\label{EE}
    \mathcal{E}:={\cal E}^{\mu}_\mu=- \frac{2 V_2}{\eta^2}+\frac{3 \Upsilon^2}{2 \Omega^4}+\frac{2}{\ell^2} -\frac{\Upsilon \partial_\phi \eta}{\Omega^3 \eta}-\frac{(\partial_\phi \eta)^2}{2\eta^2 \Omega^2} +\frac{2\mathcal{T}_{\phi \phi}}{\Omega^2}=0\, ,
\end{equation}
\begin{equation}\label{El}
    \mathcal{E}_{ll}:=l^\mu l^\nu \mathcal{E}_{\mu \nu}=\mathcal{T}_{ll}-\frac{1}{\mu \Omega} \left[\omega \mathcal{T}_{ll} -\partial_\phi \mathcal{T}_{ll}+\partial_v \mathcal{T}_{l\phi} -\partial_{\phi}(\mathcal{U} \mathcal{T}_{l\phi})- \kappa\mathcal{T}_{l\phi} + \frac{\chi \mathcal{T}_{l\phi}}{\Omega} \right]=0 \, ,
\end{equation}
\begin{equation}\label{Elp}
\begin{split}
  \mathcal{E}_{l\phi} :=q_{\phi}{}^{\mu} l^{\nu}\mathcal{E}_{\mu \nu} =\mathcal{T}_{l\phi}
    -\frac{1}{\mu \Omega} \biggl[&  \partial_v \mathcal{T}_{\phi \phi} -\mathcal{U}\partial_\phi \mathcal{T}_{\phi \phi} -2 \mathcal{T}_{\phi \phi} \partial_\phi \mathcal{U}- \frac{\chi \mathcal{T}_{\phi \phi}}{2 \Omega} \\ &
    -\Omega \partial_\phi \left( \frac{\mathcal{T}_{l\phi}}{\Omega}\right)- \omega \mathcal{T}_{l \phi} -\tau \Omega \mathcal{T}_{ll}-\frac{\chi}{2}\Omega  \mathcal{E}\biggr]=0 \, .
\end{split}
\end{equation}
As the VCT case, equations $\mathcal{E}_{ll}=0$ and $\mathcal{E}_{l\phi}=0$ can be treated as equations that determine $\kappa$ and $\mathcal{U}$. 
Unlike the VCT class, second order terms in the metric appear at leading order in the equations of motion. This is the case for \eqref{EE} where $V_2$ appears. Hence this equation cannot be used to determine a third leading order term in \eqref{line-element-001}.
 Therefore, the NVCT solution space is described by four independent functions of $v,\phi$; three of them are parametrizing the BDoF of the solution space and the last one, which may be taken to be $\mathcal{T}_{ll}$ (or $\tau$), encodes the degree of freedom associated with the massive chiral graviton. 

%%%%%%%%%%%%%%%%%%%%%%%%%%%%%%%%%%%%%%%%%%%%%%%%%%%%%%%%%%%%%%%%%%%%%
\subsection{Null boundary symmetries and surface charges}
%%%%%%%%%%%%%%%%%%%%%%%%%%%%%%%%%%%%%%%%%%%%%%%%%%%%%%%%%%%%%%%%%%%%%%%

The null boundary preserving diffeomorphisms which generate symmetries of the solution phase space are given by \eqref{xi-sym-gen}. These generate field variations \eqref{general-field-variations} over the solution phase space. The algebra of the symmetry generators are given in \eqref{3d-NBS-KV-algebra}, \eqref{W12-T12-Y12}.

To compute the charge variations associated with the symmetry generating diffeomorphisms we use the general equation in section \ref{sec:2.2}. The density of surface charge variation can be rewritten as
\begin{equation}\label{charge-density-generic-001}
    \begin{split}
         16 \pi G \mathcal{Q}^{vr}_{\text{\tiny LW}} &\approx 16 \pi G \mathcal{Q}^{vr}_{\text{\tiny GR}}+\frac{1}{\mu} \left(\delta \Gamma^{\alpha}_{\phi \beta} \nabla_\alpha \xi^\beta + \frac{1}{\ell^2}h_{\phi \alpha} \xi^\alpha \right) %\\ &
         -\frac{2}{\mu} \left(\delta \mathcal{T}_{\phi \alpha} \xi^\alpha +\xi^\beta \mathcal{T}^\alpha_{[\beta} h_{\phi]\alpha} \right)\\
       %16 \pi G \mathcal{Q}^{vr}_{\text{\tiny LW}}|_{r=0} 
      & \approx
        \hat{Y} \, \delta \left\{\tilde{\Upsilon} -\frac{2}{\mu} \left[ \mathcal{T}_{\phi \phi} +\Omega \partial_\phi\left(\frac{\mathcal{T}_{l\phi}}{\chi} \right)\right] \right\} 
     %  \\ &
       + \left[\delta_\xi \left(\Omega+\frac{1}{2\mu} \, \frac{\hat{\Upsilon}}{\Omega} \right) -\partial_\phi \left( \frac{\Omega\hat{T} \mathcal{T}_{ll}}{\mu\chi^2}\right)\right]\, \delta \mathcal{P}\\
       &-\delta \hat{\Omega} \left(\delta_\xi \mathcal{P}-\frac{2\Omega \mathcal{T}_{ll}\,\hat{T}}{ \chi^2}\right) - \frac{2\, \hat{T} \, \mathcal{T}_{ll} \, \delta \Omega}{\chi^2 } \left( \Omega - \frac{1}{\mu} \, \frac{\hat{\Upsilon}}{2 \Omega} + \frac{1}{\mu} \frac{\Omega\,\mathcal{T}_{l\phi}}{\chi}\right) \\ 
    &  +\frac{1}{\mu} \biggl[ \delta \mathcal{U}\, \delta_\xi (\tau \, \Omega) -\delta_\xi \mathcal{U}\, \delta (\tau \, \Omega) + \frac{1}{2 \Omega^2} \left( \delta_\xi \Omega \,\partial_\phi \delta \Omega - \delta \Omega \,\partial_\phi \delta_\xi \Omega\right)\\
    & \qquad  +   \frac{1}{4\eta^2} \left( \delta_\xi \eta \,\partial_\phi \delta \eta - \delta \eta \,\partial_\phi \delta_\xi \eta\right)+2 \delta \Omega \, \delta_\xi \left(\frac{\mathcal{T}_{l\phi}}{\chi} \right)-2\delta_\xi \Omega \, \delta \left(\frac{\mathcal{T}_{l\phi}}{\chi} \right)\\
    & \qquad+  \frac{\delta_\xi \Omega}{\Omega}\,\partial_\phi \delta\mathcal{P} -\frac{\delta \Omega}{\Omega}\,\partial_\phi \delta_\xi\mathcal{P}\biggr]\,.
    %\\
     %&  -\frac{2 \hat{T}}{\mu \,\chi^2 }\,\delta \Omega \left[  \partial_\phi \mathcal{T}_{ll}- \frac{2 \mathcal{T}_{ll} \partial_\phi \chi}{\chi}\right]  \\ & 
     %-\frac{\delta \mathcal{U} \, \Omega^2 \,\hat{T}}{2\mu \chi} \, \mathcal{E}+\frac{2 \hat{T}\, \Omega \delta \Omega}{\chi^2 }\, \mathcal{E}_{ll}
     \end{split}
\end{equation}
The last three lines of the above equation have been written in a suggestive way, ready to be absorbed into  the $Y$-ambiguity term
\begin{equation}\label{Y-term-NVCT}
    B_\lambda [\delta g ; g] = -\frac{1}{8} \,\Gamma^\alpha_{\lambda \beta} h^\beta_\alpha +\frac{1}{2}\, n_\alpha l^\beta \delta \Gamma^{\alpha}_{\lambda \beta} {+ {\delta \Omega} \left(  \frac{l^\alpha \mathcal{T}_{\alpha \lambda}}{\chi} - \frac{1}{2{\Omega} } \partial_\lambda \mathcal{P}\right)}\,,
\end{equation}
whose $\phi$ component is
\begin{equation}
   2 B_\phi = -\frac{\delta \Omega \partial_\phi \Omega}{2 \Omega^2}-\frac{\delta \eta \partial_\phi \eta}{4 \eta^2}+ \Omega \tau \, \delta \mathcal{U} - \delta \omega {+2\delta \Omega \, \frac{\mathcal{T}_{l\phi}}{\chi} -\frac{\delta \Omega}{\Omega} \partial_\phi \mathcal{P}} +\mathcal{O}(r)\,.
\end{equation}
Subtracting off the $Y$-term, the charge variation becomes
\begin{equation}\label{charge-NVC-case}
    \begin{split}
       \slashed{\delta}{Q}(\xi) \approx\ \frac{1}{16 \pi G} \int_{0}^{2 \pi} \d \phi \,\biggl\{ & \tilde{W}\,\delta {\tilde{\Omega}}+\tilde{Y} \, \delta \mathcal{J} +\tilde{T}\, \delta \mathcal{P}\\
     &- \frac{2\, \tilde{T} \, \mathcal{T}_{ll} \, \Omega\delta \Omega}{\chi^2 } +  \frac{2\, \Omega \, \mathcal{T}_{ll} \, \delta \Omega}{\chi^2 } \,\partial_\phi \bigg[ \tilde{Y}\big( \Omega+\frac{1}{2\mu} \, \frac{\hat{\Upsilon}}{\Omega} \big)\bigg] \biggr\}
    \end{split}
\end{equation}
with
\begin{equation}
    \mathcal{J}:=\tilde{\Upsilon} -\frac{2}{\mu} \left[ \mathcal{T}_{\phi \phi} +\Omega \partial_\phi\left(\frac{\mathcal{T}_{l\phi}}{\chi} \right) \right]\,
\end{equation}
where $\tilde{Y}=\hat{Y}$ and with the change of slicing 
\begin{align}
    \tilde{W}:=&-\delta_\xi \mathcal{P} +\frac{2 {\Omega}\hat{T}\, \mathcal{T}_{ll}}{\chi^2}= \hat{W}+ 2 \partial_\phi \hat{Y} -\hat{Y}\partial_\phi \mathcal{P} \\ \nonumber
        \tilde{T}:=&\delta_\xi \left(\Omega+\frac{1}{2\mu} \, \frac{\hat{\Upsilon}}{\Omega} \right)-\partial_\phi \left( \frac{\Omega\hat{T} \mathcal{T}_{ll}}{\mu\chi^2}\right) \\
        = \, & \frac{\hat{T}}{\Omega} \left( \Omega - \frac{1}{\mu} \, \frac{\hat{\Upsilon}}{2 \Omega} + \frac{1}{\mu} \frac{\Omega\,\mathcal{T}_{l\phi}}{\chi}\right) + \partial_\phi \left[ \hat{Y}\left( \Omega+\frac{1}{2\mu} \, \frac{\hat{\Upsilon}}{\Omega} \right)\right] \,.
        %\tilde{T}:=&\delta_\xi \left(\hat{\Omega}+\frac{1}{2\mu} \, \partial_\phi \mathcal{P} \right)-\partial_\phi \left( \frac{\Omega\hat{T} \mathcal{T}_{ll}}{\mu\chi^2}\right) \\
        %= \, & \frac{\hat{T}}{\Omega} \left( \Omega - \frac{1}{\mu} \, \frac{\hat{\Upsilon}}{2 \Omega} + \frac{1}{\mu} \frac{\Omega\,\mathcal{T}_{l\phi}}{\chi}\right) + \partial_\phi \left[ \hat{Y}\left( \hat{\Omega}+\frac{1}{2\mu} \, \partial_\phi \mathcal{P} \right)\right] 
    \end{align}
The charge variation \eqref{charge-NVC-case} can also be written as
\begin{equation}\label{charge-variation-NVCT}
    \begin{split}
       \slashed{\delta}{Q}(\xi) \approx\ \frac{1}{16 \pi G} \int_{0}^{2 \pi} \d \phi \,\biggl\{ & \tilde{W}\,\delta \mathcal{S}+\tilde{Y} \, \delta \mathcal{J} +\tilde{\mathbb{T}}\, \delta \mathcal{P}
       %\\ &
     - \frac{2\Omega\, \mathcal{T}_{ll} \, \delta \Omega}{\chi } \left( 1 -  \frac{\hat{\Upsilon}}{2\mu \Omega^2} +  \frac{\mathcal{T}_{l\phi}}{\mu\chi}\right)\, {T} \biggr\}
     % \slashed{\delta}{Q}(\xi) \approx\ \frac{1}{16 \pi G} \int_{0}^{2 \pi} \d \phi \,\biggl\{ & \tilde{W}\,\delta \tcb{\tilde{\Omega}}+\hat{Y} \, \delta \mathcal{J} +\tilde{T}\, \delta \mathcal{P}\\
     %&- \frac{2 \, \mathcal{T}_{ll} \, \delta \Omega}{\chi } \left( \Omega - \frac{1}{\mu} \, \frac{\hat{\Upsilon}}{2 \Omega} + \frac{1}{\mu} \frac{\Omega\,\mathcal{T}_{l\phi}}{\chi}\right)\, \xi^v \biggr\}
    \end{split}
\end{equation}
where ${\cal S}, {\tilde{\mathbb{T}}}$ are defined in \eqref{calS-def}, \eqref{calJ-bbT-def}.
As already stated the charge variation is not integrable which is due to the presence of flux through the boundary. However the change of slicing is essential to interpret the different elements of the surface charges and their algebra. We elaborate on these in the rest of the section.   

To treat the non-integrable charges, we use the Barnich-Troessaert modified bracket (MB) method \cite{Barnich:2011mi}. The charge variation \eqref{charge-NVC-case} is split into two, integrable and non-integrable, parts $\slashed{\delta} Q(\xi) = \delta Q^{\text{I}}(\xi) + {F}(\delta g; \xi)$ where
\begin{equation}\label{charge-NVC-case-Integrable part}
\begin{split}
       {Q}^{\text{I}}(\xi) 
       %&= \frac{1}{16 \pi G} \int_{0}^{2 \pi} \d \phi \,\left[ \tilde{W}\, \tcb{\tilde{\Omega}}+\tilde{Y} \,  \mathcal{J} +\tilde{T}\, \mathcal{P} \right]\\
       & =\frac{1}{16 \pi G} \int_{0}^{2 \pi} \d \phi \,\left[ \tilde{W}\, {{\cal S}}+\tilde{Y} \,  \mathcal{J} +\tilde{\mathbb{T}}\, \mathcal{P} \right]
          \end{split}
\end{equation}
\begin{equation}\label{charge-NVC-case-non-integrable part}
    \begin{split}
       {F}(\delta g; \xi) 
       %&= -\frac{1}{16 \pi G} \int_{0}^{2 \pi} \d \phi \,\biggl[  \tilde{T}\left(\frac{2 \, \Omega \mathcal{T}_{ll} \, \delta \Omega}{\chi^2 }\right) +  \tilde{Y}\left( \tcb{\tilde{\Omega}}+\frac{1}{2\mu} \, \partial_\phi \mathcal{P} \right)\,\partial_\phi \left(\frac{2\, \Omega \, \mathcal{T}_{ll} \, \delta \Omega}{\chi^2 } \right)  \biggr]\\
      % & =-\frac{1}{16 \pi G} \int_{0}^{2 \pi} \d \phi \,\bigg\{  \tilde{\mathbb{T}}\left(\frac{2 \, \Omega \mathcal{T}_{ll} \, \delta \Omega}{\chi^2 }\right) +  \bigg[\tilde{Y}\left( \mathcal{S}+\frac{1}{4\mu} \, \partial_\phi \mathcal{P} \right)\tcb{+}\frac1{4\mu}\tilde{W}\bigg]\,\partial_\phi \left(\frac{2\, \Omega \, \mathcal{T}_{ll} \, \delta \Omega}{\chi^2 } \right)\biggr\}  \\ 
       &= {-\frac{1}{8 \pi G} \int_{0}^{2 \pi} \d \phi \, \left(\frac{ \, \Omega \mathcal{T}_{ll} \, \delta \Omega}{\chi^2 }\right) \bigg\{ \tilde{\mathbb{T}}-  \partial_\phi\bigg[\tilde{Y}\left( \mathcal{S}+\frac{1}{4\mu} \, \partial_\phi \mathcal{P} \right){+}\frac1{4\mu}\tilde{W}\bigg]
       %\\ &- \tilde{W} \,\partial_\phi \left(\frac{ \Omega \, \mathcal{T}_{ll} \, \delta \Omega}{2\, \mu \,\chi^2 } \right)
       \bigg\}}\,,
    \end{split}
\end{equation}
where the non-integrable is associated to the flux \cite{Barnich:2011mi}. Indeed, we note that the null boundary symmetries generating vector field \eqref{xi-sym-gen} on the null surface $\mathcal{N}$ takes the form $\xi = {T} l + \hat{Y} \partial_\phi$. The flux $F$ is hence through the null surface because $l$ is perpendicular to $\mathcal{N}$. Moreover, the flux $F$ \eqref{charge-NVC-case-non-integrable part} is proportional to ${\cal T}_{ll}$ at $r=0$, which recalling \eqref{GRE} and the metric ansatz \eqref{line-element-001}, ${\cal T}_{ll}=R_{ll}$. That is the flux is proportional to the Ricci curvature along the null surface. Assuming the classical null curvature condition $R_{ll}\geq 0$ the flux is always non-positive,\footnote{For $R_{ll}\geq 0$ we have the usual focusing theorem stating that the expansion $\Theta_{_{l}}$ is never increasing in time.} in line with the usual classical intuition that the flux just passes inside (inward) through the null surface. 

The transformation laws of the integrable parts of the charges are then obtained as, 
\begin{subequations}\label{charge-transf-NVCT}\begin{align}
    \delta_\xi \mathcal{S} & = \tilde{\mathbb{T}}+ \partial_\phi \left( \frac{ {\Omega}\, \hat{T}\, \mathcal{T}_{ll}}{2 \,\mu\,\chi^2}\right) \\
    \delta_\xi \mathcal{P} &=-\tilde{W} +\frac{2 {\Omega}\,\hat{T}\, \mathcal{T}_{ll}}{\chi^2} \\\nonumber
    \delta_{\xi}\mathcal{J} &\approx \tilde{Y}\partial_{\phi}\mathcal{J}+2\mathcal{J}\partial_{\phi}\tilde{Y}-\frac{2}{\mu}\partial_{\phi}^{3}\tilde{Y} \\ &\qquad +2\hat{\Upsilon}\partial_{\phi}\left(\frac{\mathcal{T}_{ll}\hat{T}}{\mu\chi^2}\right)+\left(\frac{\mathcal{T}_{ll}\hat{T}}{\mu\chi^2}\right)\partial_{\phi}\hat{\Upsilon}-2\Omega\partial_{\phi}\left(\frac{\Omega\mathcal{T}_{ll}\mathcal{T}_{l\phi}\hat{T}}{\mu\chi^3}\right).
\end{align}
\end{subequations}
where we did not write $\hat{T}$ in terms of $\tilde{\mathbb{T}}$ and $\tilde{Y}$ for brevity. Also equations of motion for NVCT case imply that
\begin{equation}\label{EOM-for-cal-J}
    \partial_v \mathcal{J}-\mathcal{U}\partial_{\phi}\mathcal{J}-2\mathcal{J}\partial_{\phi}\mathcal{U}+\frac{2}{\mu}\partial_{\phi}^{3}\mathcal{U}-2\hat{\Upsilon}\partial_\phi \left( \frac{\mathcal{T}_{ll}}{\mu\, \chi}\right)-\frac{\mathcal{T}_{ll}}{\mu\, \chi} \partial_\phi\hat{\Upsilon} +2\Omega \partial_\phi \left( \frac{\Omega\mathcal{T}_{ll}\mathcal{T}_{l\phi}}{\mu\, \chi^2}\right) \approx 0 \, .
\end{equation}

The integrable--non-integrable split in \eqref{charge-NVC-case-Integrable part} and \eqref{charge-NVC-case-non-integrable part} was somewhat arbitrary. This arbitrariness can be fixed using the  modified bracket (MB) method \cite{Barnich:2011mi}. Here we skip the details and  interested reader can e.g. see \cite{Adami:2020amw} for a quite similar analysis. Going through the modified bracket analysis reveals that $Q^{\text{I}}(\xi), F(\delta_{\xi_{1}}g;\xi_{2})$ \eqref{charge-NVC-case-Integrable part}, \eqref{charge-NVC-case-non-integrable part} satisfy,
\begin{equation}\label{BT-Bracket-01}
    \left\{Q^{\text{I}}(\xi_{1}),Q^{\text{I}}(\xi_{2})\right\}_{\text{\tiny{MB}}} := \delta_{\xi_{2}}Q^{\text{I}}(\xi_{1})+F(\delta_{\xi_{1}}g;\xi_{2})
\end{equation}
with
\begin{equation}\label{BT-Bracket-02}
     \left\{Q^{\text{I}}(\xi_{1}),Q^{\text{I}}(\xi_{2})\right\}_{\text{\tiny{MB}}}=Q^{\text{I}}([\xi_{1},\xi_{2}]_{{\text{adj. bracket}}})+K_{\xi_{1},\xi_{2}}.
\end{equation}
Here $F$ is a field-dependent expression and $K$ is the field-independent central charge, 
\begin{equation}\label{central-charge-general}
    K_{\xi_1,\xi_2}=\frac{1}{16\pi G}\int_{0}^{2\pi}d\phi\left[\tilde{W}_{1}\tilde{\mathbb{T}}_2-\frac{1}{\mu}\tilde{Y}_{1}\partial_{\phi}^3 \tilde{Y}_{2} - (1 \leftrightarrow 2)\right].
\end{equation} 
The central term $K$ is the same as \eqref{central-extension-term} and the algebra is the same as \eqref{VCT-algebra-SPJ}. We therefore, recover the same Heisenberg $\oplus$ Virasoro algebra as in the VCT case and as explicitly seen, the flux \eqref{charge-NVC-case-non-integrable part} vanishes on-shell for the VCT case. 

We should stress that what we call $K$ is a bit different than the one discussed in \cite{Barnich:2011mi}: the central extension in \cite{Barnich:2011mi} can be field-dependent and satisfies a ``generalized 2 cocycle condition'', whereas our $K$ is field independent and satisfies usual 2 cocycle condition under the modified bracket. In other words, \eqref{BT-Bracket-02} defines an algebra with a usual central extension. Of course, as we have seen in earlier analysis in this paper, being field dependent or not depends very much on the slicing. In particular, by field-independent $K$ here we mean that there exists a slicing where $K_{\xi_1, \xi_2}$, which is  antisymmetric $K_{\xi_1, \xi_2}=-K_{\xi_2, \xi_1}$ and linear in both of $\xi_1, \xi_2$, become field independent. Expression \eqref{central-charge-general} is an example of such field-independent $K$. 

As will become more apparent in the next part, the $K$ and $F$ terms carry different physical and mathematical meanings:  presence of $K$ leads to charge non-conservations which are reminiscent of quantum anomalies, while a non-zero $F$ is due to a classical flux of charges (or bulk degrees of freedom) through the boundary. Therefore, we will exclusively call the field-independent $K$-term, the central charge and the field-dependent $F$-term, the flux.

\subsection{Chiral massive news and generalized charge conservation equation}\label{sec:GCCE}

To understand better the physical meaning of the central charge $K$ and the $F$-term flux, we  study more closely  the modified bracket equation,
\begin{equation}\label{BT-bracket-2}
    \delta_{\zeta}Q^{\text{I}}(\xi) -Q^{\text{I}}([\xi,\zeta]_{_{\text{adj. bracket}}})-K_{\xi,\zeta}\approx -F(\delta_{\xi}g;\zeta)
\end{equation}
where $\zeta, \xi$ are two arbitrary symmetry generators.  

We crucially note that \eqref{BT-bracket-2} and in particular the adjusted bracket there, is written in the tilde-slicing. As in section \ref{sec:GCCE-4.4}, let us  rewrite \eqref{BT-bracket-2} for $\zeta=\zeta (1,0,0)$ in the tilde-slicing and $\xi$ arbitrary. One can always find a local coordinate $\tilde v$ such that $\partial_{\tilde v}={\zeta}(1,0,0)$. Recalling \eqref{tilde-basis-adj-bracket}, we have $[\partial_{{\tilde v}},\xi]_{_{\text{adj. bracket}}} =0 $ for any $\xi(\tilde{\mathbb{T}}, \tilde W, \tilde Y)$ and therefore, {using the definition }
\eqref{BT-bracket-2}, \eqref{hamilevoleq} takes the form,
\begin{equation}\label{GCCE}
\begin{split}
        \frac{\d{}}{\d{}{\tilde v}} Q^{\text{I}}(\xi) := \delta_{\partial_{\tilde v}} Q^{\text{I}}(\xi) &+\frac{\partial}{\partial{\tilde v}} Q^{\text{I}}(\xi) {\approx} K_{\xi,\partial_{\tilde v}}- F(\delta_\xi g,\partial_{\tilde v}) +\frac{\partial}{\partial{\tilde v}} Q^{\text{I}}(\xi)
        ,\\
 K_{\xi,\partial_{\tilde v}}= \frac{1}{16\pi G}\int_0^{2\pi} \d \phi\ \tilde W, &\qquad       
F(\delta_\xi g,\partial_{\tilde v})  \approx\ -\frac{1}{8 \pi G} \int_{0}^{2 \pi} \d \phi 
      \frac{\Omega\, \mathcal{T}_{ll} \,}{\chi^{2} }  \delta_\xi \Omega
\end{split}
\end{equation}
where we used \eqref{central-charge-general}. {Recalling \eqref{charge-NVC-case-Integrable part}, the $\frac{\partial}{\partial{\tilde v}} Q^{\text{I}}(\xi)$ term is $Q^{\text{I}}({\partial \xi}/{\partial{\tilde v}})$  plus a term coming from $\partial_{\tilde v}$ of ${\cal S}, {\cal P}$ or ${\cal J}$. Using explicit expressions for ${\cal S}, {\cal P}$ or ${\cal J}$ in terms of fields on the solution space and once one chooses the equations of motion for boundary degrees of freedom, \eqref{GCCE} is expected to become an identity. One should, however, note that in our maximal boundary phase space setting, we do not specify dynamics of BDoF.}

Equation \eqref{GCCE} which holds for an arbitrary vector field $\xi$ is the main result of this section and reveals the physical meaning of the flux and the central terms. In the absence of genuine flux $F$ and the central term $K$, we arrive at the identity $\frac{\d{}}{\d{}{\tilde v}} Q^{\text{I}}(\xi) =\frac{\partial}{\partial{\tilde v}} Q^{\text{I}}(\xi)$. This equation, which was dubbed as Generalized Charge Conservation Equation (GCCE) in \cite{Adami:2020amw}, relates non-integrability of the charge to its $\tilde v$ dependence.  GCCE  is a generalization and extension of the ``flux-balance equation'' in the context of 4d asymptotically flat  gravity and the BMS charges, where the flux is called ``Bondi news'' \cite{Bondi:1962}, see \cite{Barnich:2011mi, Wald:1999wa} and  \cite{compere:2019gft} for some recent discussions and references.   GCCE states how the fluxes which pass through the boundary are imprinted in the corresponding surface charges. In our case the flux is associated with the chiral massive gravitons through the null boundary at $r=0$ and as we see from \eqref{GCCE}, the flux vanishes for backgrounds with ${\cal T}_{ll}=0$. This happens for all VCT backgrounds and a class of NVCT backgrounds that we discuss in the next subsection.

We should stress that the GCCE is different than similar equation in 4d flat space e.g. discussed in \cite{Adami:2020amw, Wald:1999wa,Barnich:2011mi}, because we not only have the news, the flux term $F$, but also there is a field independent central charge contribution which is absent in those analysis. {In the $3d$ Einstein gravity analysis \cite{Adami:2020ugu}, there is no flux but the Heisenberg central charge $K$ is also present and sources the non-conversation of charges. }

As a last comment, we stress again that the GCCE is written in terms of the time $\tilde v$ and not of the coordinate time $v$, used to write the metric expansion \eqref{line-element-001}. One could not have performed a similar analysis using the time $v$, since $\partial_v$ is not field-independent in the tilde-slicing, \emph{cf.} footnote \ref{v-vs-tilde v-footnote}.

%%%%%%%%%Equation%%%%%%%%n%s \%%%%%%%%%%%%%%%%%%%%%%%%%%%%%%%%%%%%%%%%
\subsection{Non-expanding backgrounds, example of warped solutions }\label{sec:non-expanding-warped}

In the non-expanding  $\chi=0$ case, as implied by \eqref{Tll}, ${\cal T}_{ll}=0$. 
Moreover, one can consistently set $\delta\chi=0$ in this sector. 
Therefore,  flux \eqref{charge-NVC-case-non-integrable part} vanishes and the charges are expected to be integrable. However, as in the non-expanding VCT case of section \ref{sec:non-expanding-case},  the non-expanding case should be studied more carefully as the change of slicing which brings us to the integrable slicing becomes singular. 

In the non-expanding VCT case of section \ref{sec:non-expanding-case},  the charge ${\cal P}$ associated with the $v$ translations generated by $T(v,\phi)$,  vanishes over the solution phase space. Therefore, it is pure gauge and it can be used to fix further the gauge. As we will show below, in the NVCT case this does not necessarily happen, unless ${\cal T}_{l\phi}$ also vanishes. In general we  then have three tower of charges.

We start the analysis by studying the non-expanding field equations.\footnote{{See section \ref{appen:A1} for an exhaustive analysis of  equations of motion for non-expanding axisymmetric case.}} 
The equation of motion \eqref{El} reduces to\footnote{If $\kappa\geq 0$, then \eqref{Tlphi-EoM} implies that $\mathcal{T}_{l\phi}$ is exponentially growing in time. We note that this is a feature of TMG not shared by $3d$ Einstein gravity.}
\begin{equation}\label{Tlphi-EoM}
    \frac{1}{\mu } \left[\partial_v \mathcal{T}_{l\phi} -\partial_{\phi}(\mathcal{U} \mathcal{T}_{l\phi})- \kappa\ \mathcal{T}_{l\phi} \right]=0 \, .
\end{equation}
The transformation law for $\mathcal{T}_{l\phi}$ is given by
\begin{equation}
    \delta_\xi \mathcal{T}_{l\phi}= \partial_v (T \mathcal{T}_{l\phi}) -\partial_\phi (\mathcal{U}T \mathcal{T}_{l\phi})+\partial_\phi(\hat{Y}\mathcal{T}_{l\phi}).
\end{equation}

The charge variation reads as
\begin{equation}
    \begin{split}
        16 \pi G \,\slashed \delta\mathcal{Q}^{v r}_{\text{\tiny TMG}}\big|_{r=0} & \approx\ \hat{W} \, \delta \left( \Omega + \frac{1}{\mu}\frac{\Upsilon}{2 \Omega} \right)+ \hat{Y} \delta\left\{  \Upsilon + \frac{1}{\mu} \left[ -2 \mathcal{T}_{\phi \phi} + \left( \frac{\Omega^2}{\ell^2}+ \frac{\Upsilon^2}{4 \Omega^2}\right)\right]\right\} \\
        &-\frac{2}{\mu} T  \left[ \delta \mathcal{T}_{l\phi} +\left(\frac{\delta \Omega}{\Omega}- \frac{\delta \eta}{2\eta}\right) \mathcal{T}_{l\phi}\right]\\
       & +\frac{1}{\mu} \biggl[ \delta \mathcal{U}\, \delta_\xi (\tau \, \Omega) -\delta_\xi \mathcal{U}\, \delta (\tau \, \Omega)  + \frac{1}{2 \Omega^2} \left( \delta_\xi \Omega \,\partial_\phi \delta \Omega - \delta \Omega \,\partial_\phi \delta_\xi \Omega\right) \\ & \qquad  
   +   \frac{1}{4\eta^2} \left( \delta_\xi \eta \,\partial_\phi \delta \eta - \delta \eta \,\partial_\phi \delta_\xi \eta\right)\biggr]\,.
   \end{split}
\end{equation}
Using the same $Y$-term as \eqref{B} without the last term, the charge variation becomes
\begin{equation}\label{non-exp-NVCT-charge}
      \slashed   \delta {Q}(\xi) \approx\ \frac{1}{16\pi G}\int_0^{2\pi} \d \phi \left\{\hat{W} \, \delta \hat{\mathcal{S}}+ \hat{Y} \delta \hat{\mathcal{J}}_{\text{\tiny{NVCT}}} -\frac{2}{\mu} T  \left[ \delta \mathcal{T}_{l\phi} +\left(\frac{\delta \Omega}{\Omega}- \frac{\delta \eta}{2\eta}\right) \mathcal{T}_{l\phi}\right]\right\} \, ,
\end{equation}
where
\begin{equation}
    \hat{\mathcal{J}}_{\text{\tiny{NVCT}}}:=\Upsilon + \frac{1}{\mu} \left(  \frac{\Omega^2}{\ell^2}+ \frac{\Upsilon^2}{4 \Omega^2}-2 \mathcal{T}_{\phi \phi}\right) \, .
\end{equation}
As we see unlike the non-expanding VCT case of subsection \ref{sec:non-expanding-case}, in the NVCT case we do not necessarily lose a tower of charge. The $T$ part of the charge variation \eqref{non-exp-NVCT-charge} implies that there are two distinct $\mathcal{T}_{l\phi}=0$ and $\mathcal{T}_{l\phi} \neq 0$ cases that we discuss below.

\paragraph{Non-vanishing $\mathcal{T}_{l\phi}$.} As discussed, we expect  the charges to be integrable in non-expanding cases. To see this explicitly, we introduce a change of slicing as $\hat{\textbf{T}}:= {\frac{1}{\mu}} T \mathcal{T}_{l\phi}$ for which the surface charge variation \eqref{non-exp-NVCT-charge} can be written as
\begin{equation}\label{Charge-variaiton-NVCT-non-expand-1}
         \delta {Q}(\xi) \approx\ \frac{1}{16\pi G}\int_0^{2\pi} \d \phi \left(\hat{W} \, \delta \hat{\mathcal{S}}+ \hat{Y} \delta \hat{\mathcal{J}}_{\text{\tiny{NVCT}}} +\hat{\textbf{T}} \,\delta \tilde{\textbf{P}}\right) \, ,
\end{equation}
where
\begin{equation}
    \tilde{\textbf{P}}:= %\frac{1}{\mu} 
    \ln{\left(\frac{\eta}{\Omega^2 \mathcal{T}_{l\phi}^2}\right)}\, .
\end{equation}
Transformation laws are
\begin{subequations}
    \begin{align}
        \delta_\xi \hat{\mathcal{ S}} \approx\ & \partial_{\phi}(\hat{Y} \hat{\mathcal{ S}})+ \frac{1}{2\mu} \partial_{\phi}\hat{W}+%\frac{
        \hat{\textbf{T}} %}{\mu}
        \,\\
        \delta_\xi \tilde{\textbf{P}} \approx\ & \hat{Y}\partial_\phi \tilde{\textbf{P}} -%\frac{1}{\mu} 
        \hat{W} - %\frac{4}{\mu} 
        4\partial_\phi \hat{Y} \, \\
        \delta_\xi \hat{\mathcal{J}}_{\text{\tiny{NVCT}}} \approx\ & \hat{Y} \partial_\phi \hat{\mathcal{J}}_{\text{\tiny{NVCT}}} + 2 \hat{\mathcal{J}}_{\text{\tiny{NVCT}}} \partial_\phi \hat{Y}+\hat{\mathcal{ S}} \partial_{\phi}\hat{W} -\hat{\textbf{T}} \partial_\phi \tilde{\textbf{P}} - %\frac{4}{\mu} 
        4\partial_\phi \hat{\textbf{T}} \,
    \end{align}
\end{subequations}
The above indicate that the charge algebra is not a direct sum of Heisenberg and Virasoro algebras. 
As in the case discussed in the end of section \ref{sec:charge-algebra-VCT}, upon another change of slicing the algebra takes the form of direct sum of Heisenberg $\oplus$ Virasoro. 

By making a change of slicing,
\begin{equation}
    \tilde{W}= \hat{W} -  \hat{Y} \partial_\phi \tilde{\textbf{P}} + 4 \partial_\phi \hat{Y}\, , \qquad \tilde{\textbf{T}}= \hat{\textbf{T}}+ \frac{1}{\mu}\partial_\phi \left[ \frac{1}{4} \hat{W} +\hat{Y} \left( \mu\hat{\mathcal{ S}}+ \frac{1}{4}\partial_\phi \tilde{\textbf{P}}\right)- \partial_\phi \hat{Y}\right] \, ,
\end{equation}
one can show that the charge variation can be written as
\begin{equation}
         \delta {Q}(\xi) \approx\ \frac{1}{16\pi G}\int_0^{2\pi} \d \phi \left(\tilde{W} \, \delta \tilde{\mathcal{S}}+ \hat{Y} \delta \tilde{\mathcal{J}}_{\text{\tiny{NVCT}}} +\tilde{\textbf{T}} \,\delta {\tilde{\textbf{P}}}\right) \, ,
\end{equation}
where charge densities are given by
\begin{equation}
    %\tilde{\textbf{P}}=\mu\,\hat{\textbf{P}} \,, \qquad 
    \tilde{\mathcal{ S}}=\hat{\mathcal{ S}}+\frac{1}{4 {\mu}}\partial_{\phi} {\tilde{\textbf{P}}}, \qquad \tilde{\mathcal{J}}_{\text{\tiny{NVCT}}}=\hat{\mathcal{J}}_{\text{\tiny{NVCT}}}+\tilde{\mathcal{ S}}\partial_{\phi} {\tilde{\textbf{P}}}+4\partial_{\phi}\tilde{\mathcal{ S}}+ {\frac{1}{\mu}}\partial_{\phi}^{2} {\tilde{\textbf{P}}}\, .
\end{equation}
Transformation laws,
\begin{subequations}
    \begin{align}
        \delta_\xi \tilde{\mathcal{ S}} \approx\ & \tilde{\textbf{T}}, \hspace{1 cm} \delta_{\xi} {\tilde{\textbf{P}}}=-{\tilde{W}},\,\\
        \delta_\xi \tilde{\mathcal{J}}_{\text{\tiny{NVCT}}} \approx\ & \hat{Y} \partial_\phi \tilde{\mathcal{J}}_{\text{\tiny{NVCT}}} + 2 \tilde{\mathcal{J}}_{\text{\tiny{NVCT}}} \partial_\phi \hat{Y}-\frac{8}{\mu}\partial_{\phi}^3\hat{Y}.
    \end{align}
\end{subequations}
imply that the algebra for the Fourier mode of the charges, \emph{cf.} \eqref{VCT-algebra-SPJ-Fourier},  upon quantisation $\{\cdot, \cdot\}\to -i[\cdot,\cdot]$, take the form
\begin{subequations}
		\begin{align}
		&[ \boldsymbol{\tilde{\mathcal{S}}}_n,\boldsymbol{\tilde{\mathcal{ S}}}_m]=0, \qquad [ \boldsymbol{\tilde{\mathcal{P}}}_n,\boldsymbol{\tilde{\mathcal{P}}}_m]=0\\
		&[ \boldsymbol{\tilde{\mathcal{ S}}}_n, \boldsymbol{\tilde{\mathcal{P}}}_m]=\frac{{i}}{8 G}\delta_{m+n,0},\\
		&[\boldsymbol{\tilde{\mathcal{J}}}^{\text{\tiny{NVCT}}}_n, \boldsymbol{\tilde{\mathcal{P}}}_m]=0,\qquad
		[\boldsymbol{\tilde{\mathcal{J}}}^{\text{\tiny{NVCT}}}_n, \boldsymbol{\tilde{\mathcal{ S}}}_m]=0 ,\\
		&[\boldsymbol{\tilde{\mathcal{J}}}^{\text{\tiny{NVCT}}}_n,\boldsymbol{\tilde{\mathcal{J}}}^{\text{\tiny{NVCT}}}_m]=(n-m)\boldsymbol{\tilde{\mathcal{J}}}^{\text{\tiny{NVCT}}}_{n+m}+\frac{1}{\mu G} n^3\delta_{n+m,0}.
		\end{align}
\end{subequations}
In this slicing we obtain a Heisenberg $\oplus$ Virasoro algebra, but the central charge of the Virasoro is 4 times that of \eqref{VCT-algebra-SPJ-Fourier}. We note, however, such  central charges do depend on the choice of $Y$-term and one can change the central charge up to a numerical factor; see  \cite{Compere:2015bca, Compere:2015mza} for some examples where this can happen. We expect similar option to exist here and upon an additional term to our $Y$-term we expect to be able to set the central charge equal to  that of \eqref{VCT-algebra-SPJ-Fourier}.

One can examine the $\mu\to \infty$ limit. Since the analysis is similar to those we presented in the previous section, we skip the details and quote the final result:  $\tilde{\textbf{P}}$ charges vanish, we lose one tower of charges and the $T$ transformation becomes a trivial gauge. We end up with a situation like in the VCT case, of course as expected. See section \ref{appen:A1} for examples of such backgrounds.

\paragraph{Vanishing $\mathcal{T}_{l\phi}$.} In this case one of the surface charges, associated with $T$ transformation,  vanishes and as in the case discussed in section \ref{sec:non-expanding-case} the corresponding generator becomes pure gauge. One can fix the gauge e.g. by setting $\eta=1, \delta_\xi\eta=0$. The non-zero charges, which are of course integrable take the form
\begin{equation}\label{Charge-variaiton-NVCT-non-expand-2}
         \delta {Q}(\xi) \approx\ \frac{1}{16\pi G}\int_0^{2\pi} \d \phi \left(\hat{W} \, \delta \hat{\mathcal{S}}+ \hat{Y} \delta \hat{\mathcal{J}}_{\text{\tiny{NVCT}}} \right) \, ,
\end{equation}
with the charge transformations,
\begin{subequations}
    \begin{align}
        \delta_\xi \hat{\mathcal{ S}} \approx\ & \partial_{\phi}(\hat{Y} \hat{\mathcal{ S}})+ \frac{1}{2\mu} \partial_{\phi}\hat{W}\,\\
        \delta_\xi \hat{\mathcal{J}}_{\text{\tiny{NVCT}}} \approx\ & \hat{Y} \partial_\phi \hat{\mathcal{J}}_{\text{\tiny{NVCT}}} + 2 \hat{\mathcal{J}}_{\text{\tiny{NVCT}}} \partial_\phi \hat{Y}+\hat{\mathcal{ S}} \partial_{\phi}\hat{W} \, .
    \end{align}
\end{subequations}
The charge algebra is exactly the same as the one for non-expanding VCT case \eqref{non-expanding-algebra-1}. The angular momentum aspect charge $\hat{\mathcal{J}}_{\text{\tiny{NVCT}}}$ reduces to $\hat{\mathcal{J}}$ for $\mathcal{T}_{\phi \phi}=0$ and hence we recover the results in section \ref{sec:non-expanding-case}. Nonetheless, $\mathcal{T}_{\phi \phi}$ need not be zero for this case. The warped TMG solutions that we discuss next, belong to this subspace of solution phase space.

\paragraph{Warped example.}
The metric describing the spacelike stretched black holes for $\nu^2>1$ is given in Schwarzschild coordinates by \cite{Anninos:2008fx}
\begin{equation}\label{warptrcoor}
\d s^2={-} N^2 \d t^2+\frac{\ell^2 \d{} \hat{r}^2}{4 N^2 R^2}+ R^2 (\d \theta+N_\theta \d t)^2
\end{equation}
where $\nu=\frac{\mu\ell}{3}$ and\footnote{The warped solution is not ``circular'' in the sense that it does not have $(t,\theta)\to (-t,-\theta)$ symmetry. This is due to the presence of the CS term in the action. Nonetheless, if we also change $\mu\to -\mu$, we get back a solution. This fact is also seen in our charges.}
\begin{equation}
    \begin{split}
      R^2 &=\frac{1}{4} \hat{r} \left[\left(\nu ^2+3\right) (r_-+r_+) {+} 4 \nu  \sqrt{\left(\nu ^2+3\right) r_- r_+}+3 \left(\nu ^2-1\right) \hat{r}\right]%\tcr{=\frac{g_{\theta \theta}}{\ell^2}}
      \\
      N^2 &=\frac{\left(\nu ^2+3\right) (\hat{r}-r_-) (\hat{r}-r_+)}{4 R^2} %\tcr{=\frac{\ell^6}{g_{rr}g_{\theta \theta}}}
      \\
      N_{\theta} &={-}\frac{2 \nu  \hat{r}{+}\sqrt{\left(\nu ^2+3\right) r_- r_+}}{2 R^2}\,.
      %\tcr{=\frac{g_{t\theta}}{g_{\theta \theta}}}
    \end{split}
\end{equation}
This is a stationary-axisymmetric black hole geometry with a Killing horizon at ${\hat{r}}=r_+$ generated by 
$$
\zeta_{_{\text{H}}}=\partial_t +\Omega_{_{\text{H}}} \partial_\theta 
$$
with  horizon angular velocity $\Omega_{_{\text{H}}}$  and surface gravity $\kappa_{_{\text{H}}}$,
\begin{equation}
\Omega_{_{\text{H}}}=\frac{2 }{2 \nu  r_+{+}\sqrt{\left(\nu ^2+3\right) r_- r_+}},\qquad \kappa_{_{\text{H}}}= \frac{(\nu^2+3)(r_+- r_-)}{2\ell (2 \nu  r_+{+}\sqrt{\left(\nu ^2+3\right) r_- r_+})}.
\end{equation}

Upon the coordinate transformation
\begin{equation}
    \d v= \d t + \frac{\ell }{2 N^2 R} \, \d{} \hat{r}\, , \qquad \d \phi =\d \theta - \frac{\ell N_{\theta}}{2 N^2 R}\, \d{} \hat{r}\, ,\qquad \d r={\frac{\ell}{2R}} \d{} \hat{r}
    %\d r=\frac{R}{2\ell} \d{} \hat{r}
\end{equation}
metric \eqref{warptrcoor} takes the form 
\begin{equation}\label{warptrcoor-GNC}
\d s^2={-} N^2 \d v^2+2 \d v \d r+ R^2 \left( \d \phi + N_\theta \d v \right)^2\,
\end{equation}
Expanding in powers of $r$, we therefore get
\begin{equation}\label{WB-001}
    \begin{split}
         \eta=1, \qquad \mathcal{U} &=- \Omega_{_{\text{H}}} \, ,\qquad \Omega=\frac{1}{\Omega_{_{\text{H}}}}\, ,\\
    \kappa= \kappa_{_{\text{H}}} \, , \qquad \omega =\frac{\Upsilon}{2\Omega}&=-\frac{(\nu - \ell \kappa_{_{\text{H}}})}{ \ell\Omega_{_{\text{H}}}},\qquad \tau=\frac{ {2} \nu-\ell\kappa_{_{\text{H}}}}{\ell \Omega_{_{\text{H}}}}
 \end{split}
\end{equation}
This is a NVCT background which has vanishing ${\cal T}_{ll}, {\cal T}_{l\phi}$ and a non-zero but constant ${\cal T}_{\phi\phi}$:
\begin{equation}
{\cal T}_{\phi\phi}=\frac{2(1-\nu^2)}{\ell^2\Omega_{_{\text{H}}}^2}\, .
\end{equation}
The charges for the warped TMG background are 
\begin{equation}
    \begin{split}
    \hat {\mathcal S}=&\frac{2\pi}{4G}\left(\frac{2\nu+\ell\kappa_{_{\text{H}}}}{3\nu\Omega_{\text{H}}}\right) \hat{W}_0(v)\, , \\
    \hat {\mathcal J}=&\frac{1}{8G} \left(\frac{\ell^2\kappa_{_{\text{H}}}^2+4\ell\nu\kappa_{_{\text{H}}}-\nu^2-3}{3\ell\nu\Omega^2_{\text{H}}}\right)\hat{Y}_0(v)\, .
 \end{split}
\end{equation}
See section \ref{appen:A1} for more examples in the ${\cal T}_{ll}=0, {\cal T}_{l\phi}=0$ class. {It is also interesting to note that $\nu=1 (\mu\ell=3)$ is special in the sense that ${\cal T}_{\phi\phi}=0$ and we are hence in the VCT sector. One may check that in this case the solution reduces to a BTZ black hole discussed in the previous section. }

\section{Axisymmetric solutions and an example with non-zero flux}\label{appen:axisymmetric-TMG-soln}
As reviewed in the introduction, solutions to TMG, unlike the Einstein gravity case, has not been completely classified. For our analysis we only need to consider solutions near an $r=0$ null surface. While for general case the equations of motion are quite cumbersome, assuming axisymmetry they become more manageable. Assuming $\partial_\phi$ to be a Killing vector, we may simply drop $\phi$ dependence in all the functions and we remain only with $v$ dependence. In this case, \eqref{Tmunu-components} simplifies to
\begin{equation}\label{Tmunu-axisymmetric}
  \mathcal{T}_{ll}= -\frac{1}{\Omega}(\partial^2_v\Omega-\kappa \partial_v\Omega), 
  \quad
    \mathcal{T}_{l\phi}= \frac{1}{\Omega} \partial_v (\Omega\omega),\quad
    \mathcal{T}_{\phi\phi}=-2\Omega (
   \partial_v  +\kappa) \, \tau -2 \left(\omega^2- \frac{\Omega^2}{\ell^2}\right) \, .
   \end{equation}
The equations of motion \eqref{El} and \eqref{Elp} take the form
\begin{subequations}\label{EoM-axisymmetric}
    \begin{align}
       \partial_v (\Omega  \mathcal{T}_{l\phi})- \kappa (\Omega  \mathcal{T}_{l\phi}) &= \Omega (\mu\Omega-\omega) \mathcal{T}_{ll}\label{EoM-axisymmetric-a}\\
       \partial_v   \mathcal{T}_{\phi\phi}- \frac{\partial_v\Omega}{2\Omega} \mathcal{T}_{\phi\phi}&=(\mu\Omega+\omega) \mathcal{T}_{l\phi}+\tau\Omega \mathcal{T}_{ll}\,.\label{EoM-axisymmetric-b}
    \end{align}
\end{subequations}
Equations \eqref{EoM-axisymmetric} can be viewed as two second order equations for four variables, $\Omega,  {\kappa}, \omega, \tau$. Note that ${\cal U}, \eta$ have dropped out of the equations. For example, one may solve $\omega, \tau$ as functions of $\Omega,  {\kappa}$ which parameterise the phase space. Moreover, as \eqref{EOM-for-cal-J} implies, for the axisymmetric case $\partial_v{\cal J}=0$ on-shell, and hence ${\cal J}={\cal J}_0=const.$ for the axisymmetric cases. This is of course expected as ${\cal J}$ denotes the angular momentum aspect charge and for the axisymmetric case it is  only allowed to have  a zero-mode, the angular momentum. 

\subsection{Vanishing flux  case}\label{appen:A1}
An interesting special case is axisymmetric ${\cal T}_{ll}=0$ solutions. 
In this case
\begin{equation}\label{Theta-kappa-axisymmetric}
 \Omega=\Omega_0=const,\quad \text{or} \quad \Omega=A \int^v \ e^{+\int^v \kappa}+\Omega_0,  
\end{equation}
where $A, \Omega_0$ are integration constants. This is a statement of the focusing theorem for this class of TMG solutions: For $A<0, \kappa>0$ and $\Omega$ is ever-decreasing.  

\paragraph{{Generic $\Omega$}.} 
One can hence rewrite \eqref{EoM-axisymmetric} as
\begin{subequations}
    \begin{align}
    &\partial_{v}^2({\Omega\omega})-\kappa\ \partial_{v}({\Omega\omega})=0\\ \label{appeqtau}
    & 5\omega^2+4\kappa\ \tau\Omega+4\Omega\partial_{v}\tau-2\tau\partial_{v}\Omega-3\frac{\Omega^2}{\ell^2}+2\mu\Omega\omega=const.
    \end{align}
\end{subequations}
They can be readily integrated as, 
\begin{subequations}\label{zero-flux-axisymmetric-soln}
    \begin{align}
        \mathcal{T}_{l\phi}&=\mathcal{T}^0_{l\phi}\frac{\Omega_0}{A}\frac{\partial_v\Omega}{\Omega}
        ,\qquad \omega= \frac{\Omega_0\omega_0}{\Omega} +\mathcal{T}^0_{l\phi}\frac{\Omega_0}{A}(1-\frac{\Omega_0}{\Omega})\label{zero-flux-axisymmetric-soln-a}\\
       \mathcal{T}_{\phi\phi}&=2\mathcal{T}^0_{l\phi}\frac{\Omega_0}{A}\left[\mu\Omega-\frac{\Omega_0}{3\Omega}-\mathcal{T}^0_{l\phi}\frac{\Omega_0}{A}\right]+ c\sqrt{\Omega}\,.
       \label{zero-flux-axisymmetric-soln-b}
    \end{align}
\end{subequations}
Equation \eqref{appeqtau} can be solved for $\tau(v)$. The solution phase space is hence completely specified by $\kappa(v), {\cal U}(v), \eta(v)$  and six constants of motion. 
Note that \eqref{zero-flux-axisymmetric-soln-a} allow solutions with non-vanishing $\mathcal{T}_{l\phi}$ and that the CS coefficient $\mu$ appears in the equation for ${\cal T}_{\phi\phi}$ (and not in that of ${\cal T}_{l\phi}$).

For this case, the surface charges \eqref{charge-NVC-case-Integrable part} are 
\begin{equation}\label{axi-charges}
        \mathcal{S}= \frac{2 \pi }{4 G} \left( \Omega+ \frac{\omega}{\mu}\right) \tilde{W}_0(v)\, , \quad
         \mathcal{J}=  \mathcal{J}_0\ \tilde{Y}_0(v)\, , \quad
         \mathcal{P}= \frac{1 }{8 G} \ln{\left(\frac{\eta }{(\partial_v\Omega)^2}\right)} \tilde{\mathbb{T}}_0(v)\,.
\end{equation}
Hence the solution space is spanned by three $v$-dependent BDoF. 

\paragraph{{Constant $\Omega$}.} 
For the constant $\Omega$ case, (\ref{EoM-axisymmetric-a}) implies $\omega=\omega_0=const.$ or $ \mathcal{T}_{l\phi}=\partial_v\omega=\mathcal{T}^0_{l\phi}\,  e^{\int^v \kappa}$. For the former case,  (\ref{EoM-axisymmetric-b}) implies $\mathcal{T}_{\phi\phi}=const$ and for the latter, and one can also simply integrate ${\cal T}_{\phi\phi}$ equation as
$2\mathcal{T}_{\phi\phi}- 2\mu\Omega \omega-\omega^2=\text{const.}$

The warped solution discussed in section \ref{sec:non-expanding-warped} corresponds to constant $\Omega, \omega, \mathcal{T}_{\phi\phi}$ and  $\chi=0$, $\mathcal{T}_{ll}=\mathcal{T}_{l\phi}=0$ case.

\subsection{Non-vanishing flux case}\label{appen:A2}
The simplest solutions with non-vanishing flux would be ${\cal T}_{ll}\neq 0, {\cal T}_{l\phi}={\cal T}_{\phi\phi}= 0$. 
However, a straightforward analysis shows that equations of motion \eqref{EoM-axisymmetric} 
do not admit such a solution. Since the analysis  when either ${\cal T}_{l\phi}$ or ${\cal T}_{\phi\phi}$ vanishes does not simplify  compared to the generic  case, when all three ${\cal T}_{ll}, {\cal T}_{l\phi}, {\cal T}_{\phi\phi}$ are non-vanishing, we consider the latter. Equations of motion \eqref{EoM-axisymmetric-a} and \eqref{EOM-for-cal-J} imply that
\begin{subequations}\label{EoM-axisymmetric-tau} 
    \begin{align}
       &\partial^2_v (\omega+\mu\Omega) +2\frac{\partial_v\Omega}{\Omega}\partial_v\omega- \kappa \partial_v (\omega+\mu\Omega  )=0 \\
       &\mathcal{J} =\mathcal{J}_0= const.
    \end{align}
\end{subequations}
The metric for this solution is described by 
\begin{equation}\label{line-element-axisymmetric}
    \d s^2= -V \d v^2 + 2 \eta \d v \d r+ {{\cal R}^2} \left( \d \phi + U \d v \right)^2\, , 
\end{equation}
with $\eta=\eta(v)$,
\begin{subequations}\label{r-expansion-3d-axisymmetric}
    \begin{align}
       V(v,r) =&  r V_1(v) +r^2 V_2(v)+\mathcal{O}(r^3) \\
       U(v,r) =& \mathcal{U}(v) + r U_1(v) %+r^2 U_2(v,\phi)
       +\mathcal{O}(r^2) \\
       {{\cal R}^2}(v,r) =& \Omega(v)^2  {+ r R_1(v)}% +r^2 h_2(v,\phi) 
       +\mathcal{O}(r^2)
    \end{align}
\end{subequations}
and 
\begin{equation}
     U_1  =-\frac{{2}\eta \omega}{\Omega^2},
\qquad   V_1=2\eta \kappa -2 {\partial_v \eta} \,,
\qquad  V_2=\frac{\eta^2}{\Omega^2}\left[{ \omega^2}+\frac{{3\Omega^2}}{\ell^2}{-2\kappa\tau\Omega-2\Omega\partial_{v}\tau} \right]\, .
\end{equation}
$\omega, \kappa$ are given in terms of $\Omega$ as given in \eqref{EoM-axisymmetric-tau}. The solution space is hence described by $\Omega(v), {\cal U}(v), \eta(v)$.

For this case, the charges \eqref{charge-NVC-case-Integrable part} are
\begin{equation}\label{axi-charges-int}
        \mathcal{S}= \frac{2 \pi }{4 G} \left( \Omega+ \frac{\omega}{\mu}\right) \tilde{W}_0(v)\, , \,\quad
         \mathcal{J}= \mathcal{J}_0 \,\tilde{Y}_0(v)\, , \quad
         \mathcal{P}= \frac{1 }{8 G} \ln{\left(\frac{\eta}{(\partial_v\Omega)^2}\right)} \tilde{\mathbb{T}}_0(v)\, 
\end{equation}
and the flux \eqref{charge-NVC-case-non-integrable part} is given by
\begin{equation}\label{axi-charge-flux}
F(\delta g ;\xi)= -\frac{1}{4G} \left( \frac{\mathcal{T}_{ll}\Omega \delta \Omega}{(\partial_v\Omega)^2} \right) \tilde{\mathbb{T}}_0(v) \, ,
\end{equation}
where ${\cal T}_{ll}$ is given in terms of $\Omega, \kappa$ in \eqref{Tmunu-axisymmetric}. {Note one can consistently take the vanishing flux limit of charges \eqref{axi-charges-int} and  recover charges of the axisymmetric vanishing flux  case \eqref{axi-charges}.}

We close this part by noting that on-shell ${\cal T}_{ll}$, and hence the flux, should scales like $1/\mu$. This is due to the fact that in $\mu\to\infty$ limit one should recover the Einstein-$\Lambda$ theory, the equations of motion of which are  ${\cal T}_{\mu\nu}=0$.

%%%%%%%%%%%%%%%%%%%%%%%%%%%%%%%%%%%%%%%%%%%%%%%%%%%%%%%%%%%%%5
\section{Discussion and Outlook}\label{sec:conclusion}
%%%%%%%%%%%%%%%%%%%%%%%%%%%%%%%%%%%%%%%%%%%%%%%%%%%%%%%%%%%%%%'

In this work we have continued studying the Null Boundary Symmetry program started in \cite{Adami:2020amw, Adami:2020ugu} for TMG. This is the first such example in which a propagating bulk degree of freedom is turned on and hence provides a good testing ground for the conjecture made in \cite{Grumiller:2020vvv}:  We found (1) three tower of charges, \emph{expansion aspect charge, entropy aspect charge and angular momentum aspect charge}, which label BDoF; (2) a non-zero flux through the null boundary associated with the chiral massive gravitons passing through the boundary, the chiral massive news. 

Our charges are generic functions over the null boundary, the null cylinder spanned by $v, \phi$. The charge algebra in the fundamental slicing has the same form at any constant $v$ slice. However, as we discussed and showed explicitly, the charge algebra depends on the slicing of the solution phase space. In the fundamental slicing we have a Heisenberg $\oplus$ Virasoro algebra \eqref{NB-algebra-SPJ-Fourier}: angular momentum aspect charge satisfy a Virasoro algebra at central charge $3/(G\mu)$ while commuting with the expansion and entropy aspect charges. The entropy aspect charge and expansion aspect charge satisfy a Heisenberg algebra with $1/(8G)$ as $\hbar$. Let us focus on the algebra of the entropy $\boldsymbol{\cal S}_0$ and the expansion $\boldsymbol{\cal P}_0$, $[\boldsymbol{\cal S}_0,\boldsymbol{\cal P}_0]=i/(8G)$, then uncertainties in them should satisfy $\Delta\boldsymbol{\cal S}_0 \Delta\boldsymbol{\cal P}_0\geq 1/(16G)$. 
On the other hand, as our generalized charge conservation equation (GCCE) shows the change in the entropy or expansion comes from passage of flux through the null surface. Quantisation of this flux then implies a quantisation on the entropy or the expansion. It would be desirable to more closely study this problem. 

The algebra has a direct sum form and is invariant under shift of $v$. These facts are slicing-dependent and there may exist slicings which these do not happen.
As an interesting algebraic question, one may classify all such slicings in which these two features are present. On the other hand, the fact that such $v$-independent slicings exist, dovetails with the ``corner symmetry'' picture \cite{Freidel:2020xyx,Freidel:2020svx,Freidel:2020ayo}. Namely, one may take the corner to be at constant $r,v$ codimension 2 surfaces. Our approach is, nonetheless, more general as it allows studying GCCE or flux-balance equations,  ``time variations'' of the charges and the flux (news). Moreover, this picture suggests that if instead of a null boundary we had computed the charges at timelike boundaries (i.e.  constant radius slices in an asymptotic AdS or flat spacetime), we should get the same charge algebra. It would be interesting to directly check this.

As discussed, the algebra of charges  depends on the slicing. % Moreover, there is the $Y$-ambiguity which affects the algebra and also  the central charge in the Virasoro part. 
However,  physical observables should be slicing-independent and field-redefinition invariant. Therefore, to extract physical observables one should classify such invariants. The key concept in this direction is the notion of ``solution phase space''. The solution phase space consists of bulk and boundary sectors which of course interact with each other through the flux-balance equation or GCCE. {In this work we did not specify boundary dynamics and allowed for generic boundary sources. Dynamics of BDoF may be formulated through  a ``refined equivalence principle'' \cite{Sheikh-Jabbari:2016lzm} which also takes into account the features and properties of the boundary.} 

In TMG in general the near null surface solution phase space is spanned by four functions of $v,\phi$, {three charges} $\boldsymbol{\cal S}_n(v), \boldsymbol{\cal P}_n(v), \boldsymbol{\cal J}_n(v)$ and the flux. The former are  to be viewed as  particular basis employed to span the boundary phase space part of this solution space and the flux, {which can be parametrised by}  Ricci scalar along the null surface $R_{ll}$, describe the bulk degree of freedom.\footnote{Metric near $r=0$ \eqref{line-element-001} involves 7 functions of $v,\phi$  and there are three equations of motion \eqref{EE}, \eqref{El} and \eqref{Elp} which relate three of them to the other four.} While our analysis about the solution phase space near a generic null surface is quite general, we should caution the reader that we have not constructed the full solutions away from this arbitrary null surface. In particular, in the TMG case, where we deal with third order differential equations of motion, the ``evolution'' away from the $r=0$ null surface need not necessarily be uniquely specified by the boundary data available at $r=0$; see \cite{Buchbinder:1992pe} for a thorough analysis on TMG and  solutions to its field equations.

TMG has two dimensionless parameters, $\mu\ell, \ell/G$ and  three particular regions in the parameter space has been of interest:
\begin{itemize}
    \item[1)] 
Einstein-$\Lambda$ gravity limit, $\mu\ell\to \infty$ keeping $\ell/G$ fixed. In this case, as discussed our analysis reproduces results of \cite{Adami:2020ugu}.   
\item[2)] Chiral gravity point,  $\mu\, \ell=1$ \cite{Li:2008dq}. This is special in some different ways: the massive graviton becomes massless, one of the Brown-Henneaux type central charges ($c_R$ in the conventions of section \ref{sec:GCCE-4.4}) vanishes. 
%It is known that TMG in the first order formulation can be written as $SL(2,R)_L\times SL(2,R)_R$ Chern-Simons theory at levels $c_L, c_R$. At the chiral point one of the central charges vanishes \sout{and hence the action will only involve the left sector}. For this reasons the theory at chiral point has been dubbed as chiral gravity .  
and the third order TMG field equations  linearised around AdS$_3$ background 
``degerates'' allowing for a log-mode \cite{Grumiller:2008qz,Grumiller:2008es,Carlip:2008jk,Henneaux:2009pw,Maloney:2009ck, Gaberdiel:2010xv, Grumiller:2013at}. It is hence interesting to study our charge analysis at the chiral point. We did this in part for the VCT case in section \ref{sec:non-expanding-case} (see footnote \ref{footnote12}). For the NVCT case, however, $\mu\ell=1$ does not seem to yield special features in the charges or algebra. It would, however, be interesting to study this case more thoroughly. 
 \item[3)] Conformal gravity corresponds to $\mu\ell\to 0$, keeping $\mu G$ fixed \cite{Afshar:2011yh, Afshar:2011qw}. In particular it was used in \cite{Bagchi:2012yk} as a first evidence for a holographic correspondence between asymptotically flat spacetimes and an unitary field theory (a chiral conformal field theory). The asymptotic symmetry group in \cite{Bagchi:2012yk} is precisely \eqref{VCT-algebra-SPJ-Fourier} in the conformal gravity limit. This is an indication that our results hold for asymptotic boundaries.\footnote{The first check that our program is applicable to asymptotic boundaries was done in \cite{Ruzziconi:2020wrb}.}
 \end{itemize}

The most general boundary conditions  for pure Einstein gravity with or without cosmological constant, in metric and first order (Chern-Simons) formulations have been studied in \cite{Grumiller:2016pqb,Grumiller:2017sjh} where it was discussed that the BDoF are given by six codimension two functions. This is different from our results here and in \cite{Adami:2020ugu} where we find three codimension one BDoF.  It is important to understand precisely how these results are related. Moreover, it would be interesting to build a Chern-Simon counting of the most general boundary degrees of freedom for TMG in the spirit of \cite{Grumiller:2016pqb,Grumiller:2017sjh}, and  connect with our results here.

Gravity may be formulated in the vielbein framework. % where in addition to diffeomorphisms,  the theory is invariant under local Lorentz transformations. The latter may be fixed to preserve the vielbein using Kosmann-Lie derivative  \cite{Jacobson:2015uqa,Prabhu:2015vua}. 
The covariant space formalism can be used to compute the variation of charges associated to these symmetries. However, due to the $Y$-ambiguity the results are not expected to match with the metric derivation 
%, but rather to differ by $Y$-ambiguity term 
\cite{DePaoli:2018erh,Oliveri:2019gvm,Margalef-Bentabol:2020teu,G.:2021qiz}. It would be interesting to compare the charges in the  metric and dreibein formulation of Einstein-$\Lambda$ or TMG. %\cite{Mielke:1991nn,Baekler:1991cx,Geiller:2020edh}. 
It would also be interesting to discuss the $3d$ dual charges defined in \cite{Geiller:2020edh,Geiller:2020okp} and understand whether and/or how they are included in the maximal phase space discussed in this work. 

%%%%%%%%%%%%%%%%%%%%%%%%%%%%%%%%%%%%%%%%%%%%%%%%%%%%%%%%%%%%%%'
\section*{Acknowledgement}
We would like to thank especially Daniel Grumiller for many discussions on the same line of projects and Romain Ruzziconi and Wei Song for discussions or comments. VT also would like to thank Mohammad Hassan Vahidinia for the useful discussions. MMShJ would like to acknowledge SarAmadan grant No. ISEF/M/99131. The work of VT is partially supported by IPM funds. 
CZ was supported by the Austrian Science Fund (FWF), projects P 30822 and M 2665. 

\addcontentsline{toc}{section}{References}

\providecommand{\href}[2]{#2}\begingroup\raggedright\endgroup

\end{document}